\documentclass[twocolumn,secnumarabic,amssymb, nobibnotes, aps, prd,superscriptaddress]{revtex4-1}
\usepackage{lipsum}  
\usepackage{graphics}  
\usepackage{graphicx}

\begin{document}

	\begin{abstract}
		We demonstrate a new fabrication process for hybrid semiconductor-superconductor heterostructures based on anodic oxidation (AO), allowing controlled thinning of epitaxial Al films. 
		Structural and transport studies of oxidized epitaxial Al films grown on insulating GaAs substrates reveal spatial non-uniformity and enhanced critical temperature and magnetic fields. 
		Oxidation of epitaxial Al on hybrid InAs heterostructures with a conducting quantum well show similarly enhanced superconducting properties transferred to the two-dimensional electron gas (2DEG) by proximity effect, with critical perpendicular magnetic fields up to $3.5$ T. 
		An insulating AlOx film, that passivates the heterostructure from exposure to air, is obtained by complete oxidation of the Al. It simultaneously removes the need to strip Al which damages the underlying semiconductor.
		AO passivation yielded 2DEG mobilities two times higher than similar devices with Al removed by wet etching. An AO-passivated Hall bar showed quantum Hall features emerging at a transverse field of 2.5 T, below the critical transverse field of thinned films, eventually allowing transparent coupling of quantum Hall effect and superconductivity. AO thinning and passivation are compatible with standard lithographic techniques, giving lateral resolution below $<$50 nm. We demonstrate local patterning of AO by realizing a semiconductor-based Josephson junction operating up to 0.3~T perpendicular. 			
	\end{abstract}

	\title{Anodic Oxidation of Epitaxial Superconductor-Semiconductor Hybrids}%
		
	\author{Asbj\o rn C. C. Drachmann}%

	\email{asbjorn.drachmann@nbi.ku.dk }
	\affiliation{Center for Quantum Devices and Microsoft Quantum Lab Copenhagen, Niels Bohr Institute, University of Copenhagen, Universitetsparken 5, 2100 Copenhagen, Denmark}
	
	\author{Rosa E. Diaz}%
	\affiliation{Birck Nanotechnology Center, Purdue University, West Lafayette, Indiana 47907, USA}
	
	\author{Candice Thomas}%
	\affiliation{Birck Nanotechnology Center, Purdue University, West Lafayette, Indiana 47907, USA}
	\affiliation{Department of Physics and Astronomy and Station Q Purdue, Purdue University, West Lafayette, Indiana 47907, USA}

	\author{Henri J. Suominen}%
	\affiliation{Center for Quantum Devices and Microsoft Quantum Lab Copenhagen, Niels Bohr Institute, University of Copenhagen, Universitetsparken 5, 2100 Copenhagen, Denmark}
	
	\author{Alexander M. Whiticar}%
	\affiliation{Center for Quantum Devices and Microsoft Quantum Lab Copenhagen, Niels Bohr Institute, University of Copenhagen, Universitetsparken 5, 2100 Copenhagen, Denmark}

	\author{Antonio Fornieri}%
	\affiliation{Center for Quantum Devices and Microsoft Quantum Lab Copenhagen, Niels Bohr Institute, University of Copenhagen, Universitetsparken 5, 2100 Copenhagen, Denmark}
	
	\author{Sergei Gronin}%
	\affiliation{Birck Nanotechnology Center, Purdue University, West Lafayette, Indiana 47907, USA}
	\affiliation{Department of Physics and Astronomy and Station Q Purdue, Purdue University, West Lafayette, Indiana 47907, USA}
	\affiliation{Microsoft Quantum Purdue, Purdue University, West Lafayette, Indiana 47907, USA}

	\author{Tiantian Wang}%
	\affiliation{Birck Nanotechnology Center, Purdue University, West Lafayette, Indiana 47907, USA}
	\affiliation{Department of Physics and Astronomy and Station Q Purdue, Purdue University, West Lafayette, Indiana 47907, USA}

	\author{Geoffrey C. Gardner}%
	\affiliation{Birck Nanotechnology Center, Purdue University, West Lafayette, Indiana 47907, USA}
	\affiliation{Department of Physics and Astronomy and Station Q Purdue, Purdue University, West Lafayette, Indiana 47907, USA}
	\affiliation{Microsoft Quantum Purdue, Purdue University, West Lafayette, Indiana 47907, USA}
	\affiliation{School of Materials Engineering, Purdue University, West Lafayette, Indiana 47907, USA}
	
	\author{Alex R. Hamilton}%
	\affiliation{Center for Quantum Devices and Microsoft Quantum Lab Copenhagen, Niels Bohr Institute, University of Copenhagen, Universitetsparken 5, 2100 Copenhagen, Denmark}
	\affiliation{School of Physics, University of New South Wales, Sydney, New South Wales 2052, Australia}
	\affiliation{Australian Research Centre of Excellence in Future Low Energy Electronics Technologies,  University of New South Wales, Sydney, New South Wales 2052, Australia}

	\author{Fabrizio Nichele}%
	\affiliation{Center for Quantum Devices and Microsoft Quantum Lab Copenhagen, Niels Bohr Institute, University of Copenhagen, Universitetsparken 5, 2100 Copenhagen, Denmark}
	\affiliation{IBM Research—Zurich, Säumerstrasse 4, 8803 Rüschlikon, Switzerland}
	
	\author{Michael J. Manfra}%
	\affiliation{Birck Nanotechnology Center, Purdue University, West Lafayette, Indiana 47907, USA}
	\affiliation{Department of Physics and Astronomy and Station Q Purdue, Purdue University, West Lafayette, Indiana 47907, USA}
	\affiliation{Microsoft Quantum Purdue, Purdue University, West Lafayette, Indiana 47907, USA}
	\affiliation{School of Materials Engineering, Purdue University, West Lafayette, Indiana 47907, USA}
	\affiliation{School of Electrical and Computer Engineering, Purdue University, West Lafayette, Indiana 47907, USA}
	
	\author{Charles M. Marcus}%
	\email{chmarcus@microsoft.com}
	\affiliation{Center for Quantum Devices and Microsoft Quantum Lab Copenhagen, Niels Bohr Institute, University of Copenhagen, Universitetsparken 5, 2100 Copenhagen, Denmark}
	
	\date{\today}%
	\maketitle

	\section{Introduction}
	\label{intro}

	The recent emergence of epitaxial semiconductor-superconducting hybrid materials has provided numerous experimental systems with highly controllable electronic and superconducting  properties \cite{Krogstrup2015a,Chang2015a,Wan2015,Amet2016,Shabani2016a,Kjaergaard2016,Lee2017,He2017,Krizek2018,Vaitiekenas2018,Guiducci2019,Pendharkar2019,Carrad2020,Kanne2020}.
	Interest has been sparked by theoretical predictions of topological phases in one-dimensional (1D) superconductor-semiconductor hybrids \cite{Lutchyn2010,Oreg2010} with possible application toward topological quantum computing \cite{Karzig2017}. 
	Alternatively, topological phases can be achieved in two-dimensional (2D) systems combining superconductivity with top-down fabricated narrow wires \cite{Nichele2017,OFarrell2018,Whiticar2020}, Josephson junctions \cite{Hell2017, Pientka2017, Fornieri2019} or quantum Hall edges in fully 2D systems proximitized by superconductivity \cite{Qi2010, Mong2014}. 
	
In each of these instances, the subgap density of states is a key property of the hybrid system \cite{Cheng2012,Rainis2012}. Low subgap density of states, typically extracted from the subgap conductance in a tunneling measurement, is associated with high transparency of the semiconductor-superconductor interface, that is, high Andreev reflection probability \cite{BTK, beenakker1992}.
	Near-unity interface transmission probabilities have been achieved by \textit{in-situ} molecular beam growth of epitaxial Al/InAs hybrid heterostructures in different geometries:
	Vapor-liquid-solid nanowires \cite{Krogstrup2015a,Chang2015a}, selective area grown nanowires \cite{Krizek2018,Vaitiekenas2018}, and shallow 2DEGs \cite{Shabani2016a,Kjaergaard2016,Kjaergaard2017}.  
	Recently, similar high interface transparencies were reported for other hybrid nanowires such as Sn-InSb \cite{Pendharkar2019}, Ta-InAs \cite{Carrad2020} and Pb-InAs \cite{Kanne2020}. 	
	
	Despite their high Andreev reflection probability and resulting hard induced gap, InAs/Al hybrid platforms face two important challenges: (i) surface scattering and reduced mobility resulting from the  quantum well being close to the surface, and (ii) limited critical magnetic fields due to the use of Al.  Regarding surface scattering, by design the semiconductor wave function must extend to the surface when Al is present.  Once the Al is removed, carriers are susceptible to surface scattering, resulting in a drop in mobility \cite{Hatke2017}. Sensitivity to surface chemistry of shallow 2DEGs for similar materials was recently investigated in Ref.~\cite{Pauka2019}. For hybrid heterostructures with Al, a significant contribution to surface scattering is added by the standard Al etching process \cite{Shabani2016a}. 
	Regarding low critical magnetic field, we note that magnetic fields are used in these materials for a variety of purposes, particularly to drive the topological transition \cite{Lutchyn2010,Oreg2010,Qi2010, Mong2014}. The applied superconductor should have critical fields higher that the field needed to drive the transition.
	In typical applications of 1D hybrids, a magnetic field, $B_\parallel$, along the axis of the 1D nanowire lowers states by Zeeman splitting into the superconducting gap, where, in the presence of spin-orbit coupling, a topological superconducting phase can appear. In typical applications of 2D hybrids, a perpendicular magnetic field, $B_\perp$, is applied to drive the system into the quantum Hall regime creating edge states that coexist with superconductivity. 
		
	High critical fields are found in type-II superconductors such as MoRe, with $B_{c2}\sim8\text{ T}$ \cite{Amet2016}, and NbN with $B_{c2}\sim25\text{ T}$ \cite{Lee2017}. These materials allow penetration of flux, creating vortices, that decrease Andreev reflection probability when used for proximity effect \cite{Lee2017}, potentially leading to unwanted subgap state. MoRe and NbN have been  deposited \textit{ex situ} on the edge of high mobility quantum well heterostructures such as GaAs/AlGaAs \cite{Wan2015} or Graphene/hBN \cite{Amet2016,Lee2017}, yielding superconducting features within the integer Quantum Hall regime. The studies show moderate Andreev reflection probability, possibly associated with \textit{ex-situ} deposition. 
	
	To date, superconductors showing high-transparency interfaces with semiconductors, including Al, Sn, Ta and Pb, have been type I, with relatively low critical magnetic fields, $B_c<100$ mT in bulk. However, as thin films these same materials show large in-plane critical field, $B_{c,||}$, for instance $B_{c,||}>8.5 \text{ T}$ reported for a 15~nm thick layer of Pb \cite{Kanne2020}, and $B_{c,||}>3 \text{ T}$ for 5-7~nm Al films \cite{OFarrell2018}.  Though not coupled to a heterostructure, $B_{c,||}>9 \text{ T}$ was measured \cite{Tedrow1982} on a 2~nm thin Al film and a perpendicular critical field $B_{c,\perp}\sim3.6 \text{ T}$ was extrapolated \cite{Chui1981} from granular aluminum \cite{Cohen1968, Deutscher1977}.
		
	In this work, we develop a method to apply \textit{ex-situ} anodic oxidation (AO) to epitaxial Al, the top layer of a  heterostructure stack grown by molecular beam epitaxy on III/V substrates. AO is demonstrated to thin the metallic Al, improving its material properties for these applications, and, when AO is applied with higher voltage, to fully oxidize the Al, it results in an insulating AlOx layer and a semiconductor interface that has never been exposed to air. In both applications, local patterning using electron beam lithography is employed. 
	
	The rest of the paper is organized as follows: Section \ref{AO} gives experimental details of the AO process. Section \ref{TEM} presents structural analysis of anodized Al on insulating GaAs. Section \ref{GaAs} presents transport properties of Al on insulating GaAs. Turning to InAs heterostructures with conducting quantum wells, Section \ref{InAs} presents enhanced superconducting properties after thinning of Al by AO. Passivation of the 2DEG by complete oxidation of Al is presented in Section \ref{IIIV}, followed by quantum Hall measurements on an AO-passivated Hall bar. Finally, in Section \ref{SNS}, a fabrication scheme is presented that provides high-resolution AO patterning.

	\section{Fabrication and measurement}
	\label{AO}

	\begin{figure}
		\includegraphics[width=\linewidth]{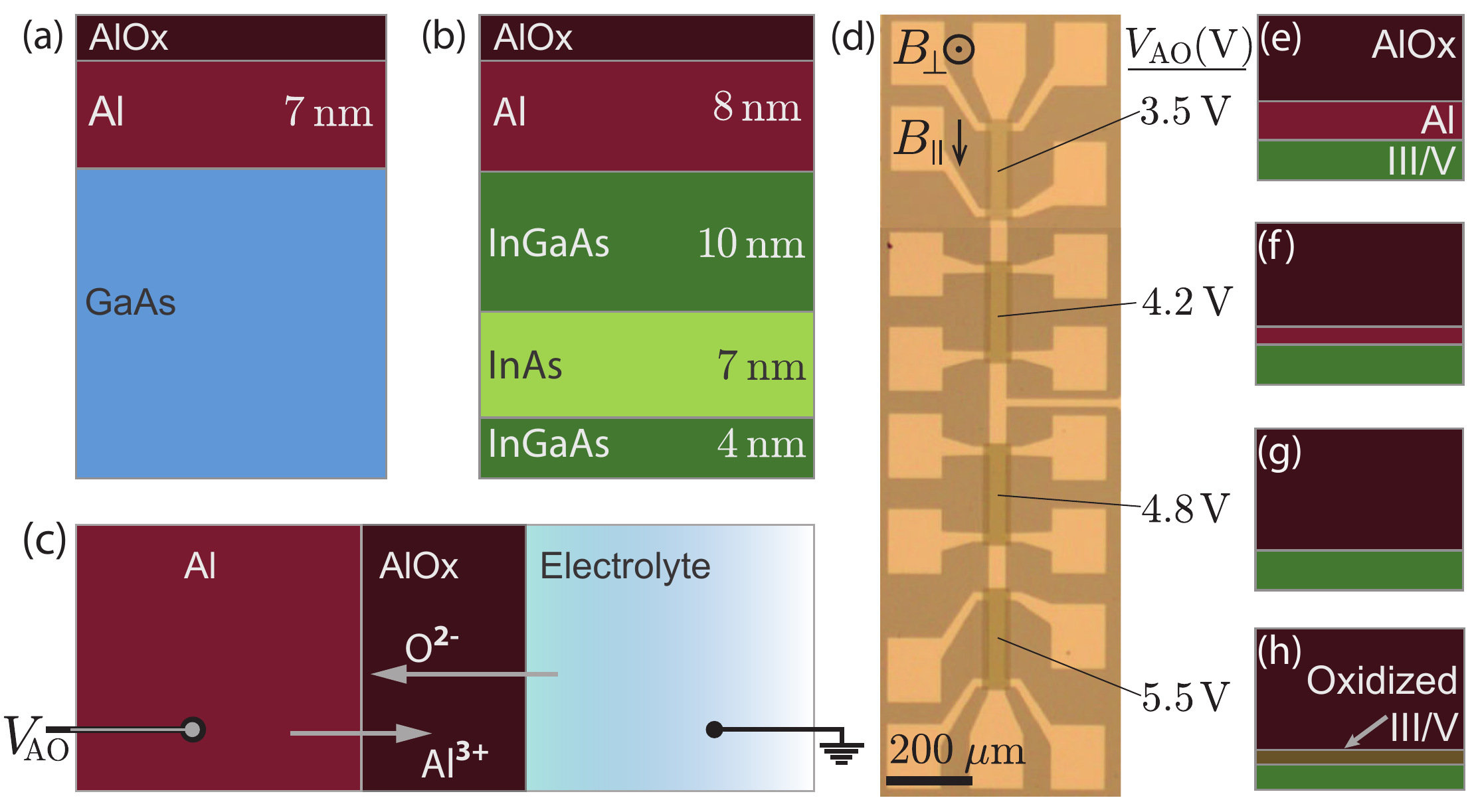}
		
		\caption{Hybrid heterostructures grown by molecular beam epitaxy, with Al films on III/V substrates. 
			\textbf{(a)} Sample 1, Al on insulating GaAs (100) substrate. \textbf{(b)}  Sample 2, Al capping a shallow InAs heterostructure supporting a two dimensional electron gas.
			\textbf{(c)} Anodic oxidation with source voltage $V_\mathrm{AO}$ between the metal (Al) and the electrolyte to drive O$^{2-}$ ions to the Al surface. 
			\textbf{(d)} Optical micrograph of an Al/InAs Hall bar (device $\alpha$) characterized by four different regions oxidized with different AO voltages, $V_\mathrm{AO}$. 
			\textbf{(e-h)} Regimes of anodic oxidation for $V_\mathrm{AO}$ between 3.5~V and 5.5~V. (e-g) Increasing $V_\mathrm{AO}$ increases oxidation depth, until only alumina is left, as indicated. (h) Further increasing $V_\mathrm{AO}$ oxidizes the underlying III/V.
		}
		
		\label{fig1}
		
	\end{figure}

Two heterostructures, denoted samples 1 and 2, were grown by molecular beam epitaxy. Sample 1 consisted of a 7~nm epitaxial Al layer grown on an {\it insulating} GaAs substrate, see Fig.~\ref{fig1}(a). This sample allowed study of the Al film as the only conductor. Sample 2 consists of an 8~nm epitaxial Al layer grown on a shallow InAs heterostructure, see Fig.~\ref{fig1}(b). Sample 2 allows the effects of AO on Al coupled to an InAs quantum well 2D electron gas (2DEG) to be investigated, including superconducting proximity effect for thin Al, and surface passivation for fully oxidized Al.

	The AO technique we employ is widely used in industry \cite{Diggle1969, Henley1982}. 
	Under atmospheric conditions, Al forms a native oxide saturating at $\sim3$~nm \cite{Hieke2017}. AO enables a deeper oxidation, with a depth controlled by an applied voltage $V_\mathrm{AO}$.
	Fig.~\ref{fig1}(c) sketches the AO setup: $V_\mathrm{AO}$ is applied between the Al and an electrolyte solution in which the substrate is submerged. The voltage drives ions to the Al/AlOx surface, increasing the oxidation depth.
	Depending on the electrolyte used, two oxide morphologies can form \cite{Henley1982}. 
	The one of interest is thin and uniform, with $V_\mathrm{AO}$ controlling the oxidation depth. This oxide type, described in Refs.~\cite{Henley1982, Nakamura1996}, can be achieved by implementing tartaric acid, 3\% by mass, as electrolyte, regulated to pH$\sim 5.5$ by ammonium hydroxide.
	A voltage-dependent oxide thickness of 1.3~$\mathrm{nm/V}$ is expected. Further details of the setup are given in  Sec.~A of Supplementary Material (SM) \cite{SM}. Standard electron beam lithography was used to pattern elongated Hall bars from an InAs 2DEG heterostructure, shown in Fig.~\ref{fig1}(d) and from Al on GaAs, in SM Sec.~B \cite{SM}. 
	
	On sample 1 [Fig.~\ref{fig1}(a)], AO was patterned into areas with $V_\mathrm{AO}=4\text{ V}$. Two areas on the sample, designated $AO_I$ and $AO_{II}$, were exposed simultaneously to check for consistency. A subsequent wet etch defined elongated Hall bars in both anodized and un-processed epitaxial Al. 
	
	On sample 2 [Fig.~\ref{fig1}(b)], grown on an InP substrate, a standard III/V wet mesa etch defined Hall bars and bonding pads for ohmic contacts. In four sequential lithography steps, AO was performed at voltages $V_\mathrm{AO} =$ 3.5~V, 4.2~V, 4.8~V, and 5.5~V on different areas, as shown in Fig.~\ref{fig1}(d). Following AO, an 18~nm HfO layer was deposited globally by atomic layer deposition, followed by deposition of patterned Ti/Au gates. Two Hall bars, denoted $\alpha$ and $\beta$, each with four AO exposures, were fabricated simultaneously on the same chip, again to check consistency. Fig.~\ref{fig1}(d) shows Hall bar $\alpha$. Further InAs fabrication details are given in SM Sec.~C \cite{SM}. These voltages yielded outcomes of two types: For smaller $V_\mathrm{AO}$, the thickness of Al was reduced and its morphology altered, but it remained a metallic conductor. For larger $V_\mathrm{AO}$, complete oxidation occurred, as illustrated in Fig.~\ref{fig1}(g). Increasing $V_\mathrm{AO}$ further began to oxidize the semiconductor, leading to increased disorder, as indicated in Fig.~\ref{fig1}(h). 
	
	Measurements were carried out in various dilution refrigerators, each with a base temperature $\sim$20~mK and a 6-1-1 T vector magnet. Standard ac lock-in techniques were used to apply a small AC current $I$ in the range 1-10~nA while measuring the longitudinal voltage, $V_{xx}$. Longitudinal resistivity $\rho_{xx}$ was obtained from the measured differential resistance, $dV_{xx}/dI$, by division with 5.0, the ratio of voltage-probe separation to Hall bar width.


	\section{Anodized A\lowercase{l} on insulating G\lowercase{a}A\lowercase{s}: \\ Structural Analysis}

	\label{TEM}

	\begin{figure}
		\includegraphics[width=\linewidth]{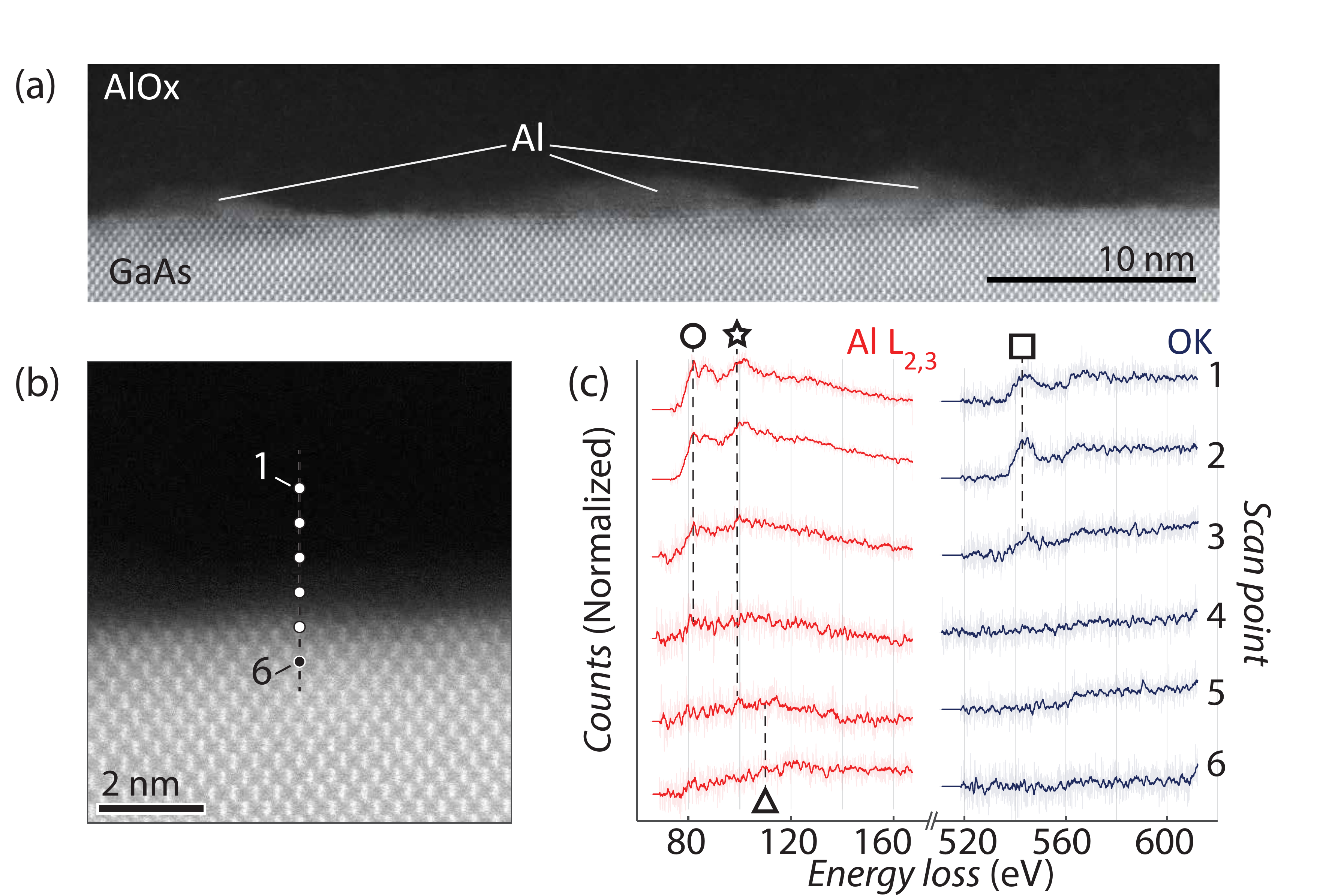}
		
		\caption{Transmission electron microscopy study of Al anodized at 4 V on GaAs substrate.
			\textbf{(a)} Scanning transmission electron micrograph in high-angle annular dark-field (HAADF) mode of $AO_I$, showing up to 2~nm hills of Al at the GaAs interface. 
			\textbf{(b)} HAADF image of the Al/GaAs interface between Al hills. Electron energy loss spectroscopy (EELS) was performed at 6 different locations (marked by points and separated by 0.3~nm) along the growth direction to gauge material composition.
			\textbf{(c)}: EELS image showing L edge for Al spectra (Red) and K edge for O (Blue) from scan points 1--6, in (b) with 0.3~nm separation. Symbols indicate composition: ($\circ$) AlOx, ($\star$) Al or AlOx, ($\triangle$) GaAs, ($\Box$) oxygen.
		}
		
		\label{fig2}
		
	\end{figure}

	To check the uniformity of Al films following AO, sample 1 was characterized with scanning transmission electron microscopy (STEM) and electron energy loss spectroscopy (EELS). 
	STEM sample was prepared (Thermo Fisher dual beam SEM/FIB Helios 4G) with accelerating voltage 500 V for final thinning of the focused ion beam (FIB) lamella.
	STEM imaging (Thermo Fisher aberration-corrected Themis Z) used an accelerating voltage of 300 kV, giving a resolution of 65 pm. STEM images were acquired along the [110] zone axis in the semiconductor. EELS data acquisition was performed in STEM mode at 300~kV with an energy resolution of 0.75 eV and a dispersion of 0.05 eV.

	Figure~\ref{fig2}(a) shows a high-angle annular dark-field (HAADF) micrograph of $AO_I$ at the GaAs/Al interface. Three Al hills (dark grey) are distinguishable from GaAs (bright) and AlOx (black).
	Hills of Al, $\sim2$~nm in height, were observed across the entire cross section. 
	Fast Fourier transform of the hills extracts a crystal growth direction along [110], see SM	Sec.~D \cite{SM}. 
	The origin of the Al hills is not known.
	Possible explanations include roughness after growth, increased oxidation at grain boundaries or degradation of the lamella during FIB preparation.
	
	EELS was implemented to analyze the elemental configuration \cite{Ahn2004,Qian2014}. Specifically, a study was done to conclude whether the region in between Al hills had metallic Al or only alumina.
	High resolution STEM micrographs in HAADF mode were acquired at the Al/GaAs interface between Al hills, one is shown in Fig.~\ref{fig2}(b). 
	EELS analysis was done at points on a line along the growth direction, see Figs.~\ref{fig2}(b,c).  Dual EELS mode was used, allowing simultaneous acquisition of the Al  L-edge ($\sim$80 eV) and O K-edge ($\sim$540 eV). All EELS data had a background subtracted, was then normalized and averaged over 20 points.
	
	When GaAs is imaged, a shoulder around 110 eV [marked by $\triangle$ in Fig.~\ref{fig2}(c)] is expected. Metallic aluminum has one peak around 97 eV ($\star$) while alumina has two peaks around 79 eV ($\circ$) and 98 eV ($\star$).  Oxygen produces a peak around 540 eV ($\square$).  At the Al L-edge, GaAs ($\triangle$) is evident at points 6 and possibly 5; alumina ($\circ$) is evident at points 1 to 3, possibly 4; metallic Al ($\star$, no $\circ$) is evident at points 4 and 5. Inspecting the O K-edge, oxygen ($\square$) is only present in points 1--3, indicating that point 4 is metallic Al. 
	
	To obtain statistics on the global morphology of the Al after AO, nine additional EELS analyses were conducted in between other Al hills. One region out of 10 failed to show metallic Al, suggesting that roughly 10\% of regions between hills are fully oxidized.
	A collection of broad-view STEM [Fig.~\ref{fig2}(a)] were used to extract the fraction of hills to off-hill regions across the film. Hills have an average height of 2~nm and cover roughly one third of the imaged regions. Within the remaining two thirds, $\sim$10\% appears fully oxidized. The remaining $2/3~\times~9/10~=~3/5$ contains a thin metallic Al layer.
	Measurements of the metallic Al thickness between hills span 3 to 6 $\mathrm{\AA}$ (5 $\mathrm{\AA}$ on average), corresponding to roughly 1-2 monolayers (ML) of Al. Additional scans are shown in SM Sec.~D \cite{SM}. In summary, an uneven, though continuous, Al film is formed by AO.

	
	\section{Anodized A\lowercase{l} on insulating G\lowercase{a}A\lowercase{s}: \\ Transport Studies}

	\label{GaAs}

	\begin{figure}
		\includegraphics[width=\linewidth]{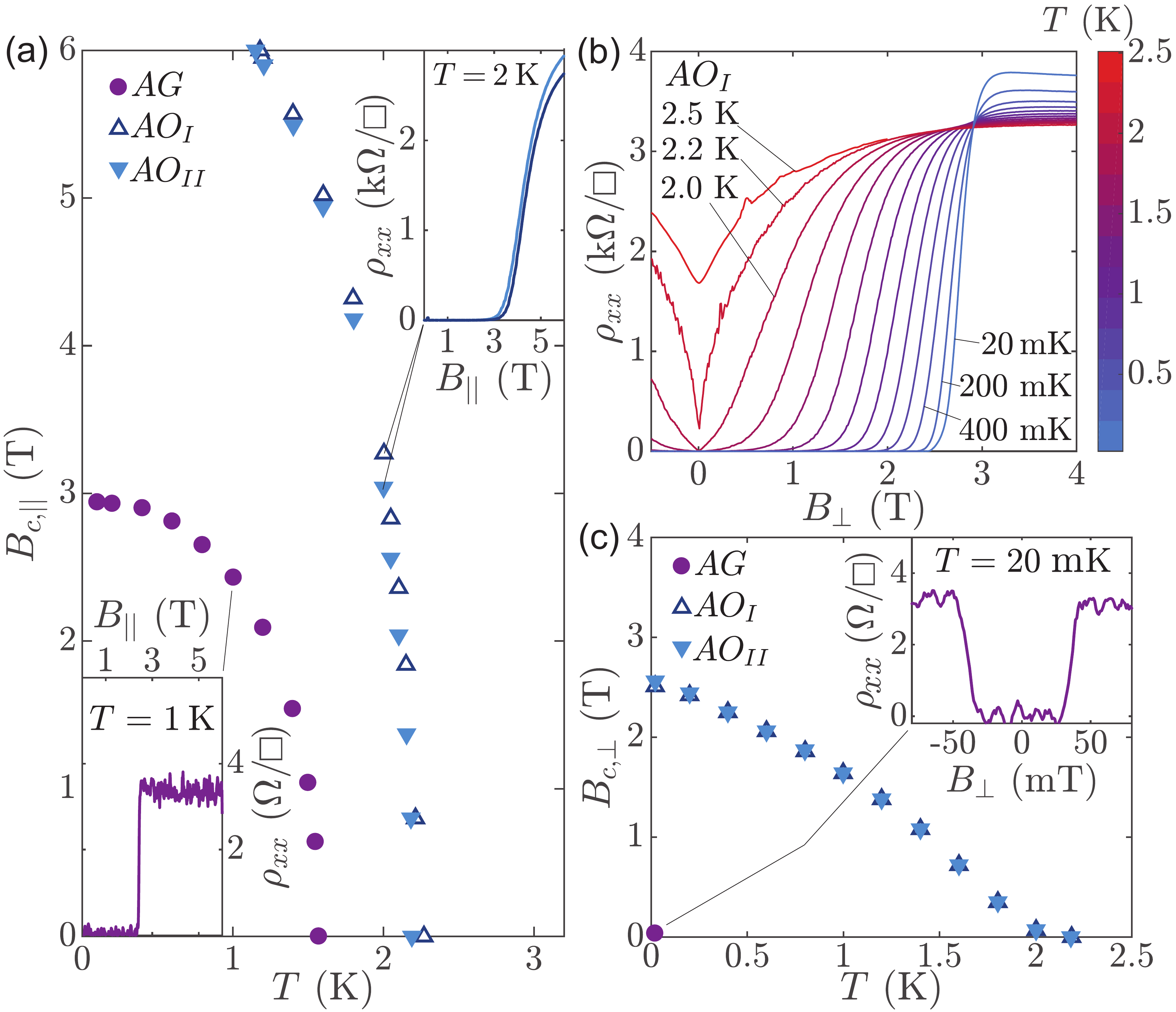}
		
		\caption{Transport data from sample 1.
			\textbf{(a)} Critical in-plane field of as-grown ($AG$) and anodic-oxidation ($AO$) Hall bars as a function of temperature, $T$. Insets: Examples of $\rho_{xx}$ versus $B_\parallel$ at elevated temperatures, 1 K and 2 K, used for extracting $B_{c,||}$ points marked by lines.
			\textbf{(b)} $AO_I$ longitudinal resistance $\rho_{xx}$ as a function of perpendicular field, $B_\perp$, at different temperatures. Note field-driven superconductor-insulator transition at $B_{\perp}\sim 3$~T.
			\textbf{(c)} Same as (a) but for $B_\perp$, extracted from data in (b) with $\rho_{xx}(B_{c,\perp})=0.01\,\rho_{N}$. Inset: AG $\rho_{xx}$ versus $B_\perp$ at base temperature.
		}
		
		\label{fig3}
		
	\end{figure}

	We investigate the effect of AO on the superconducting properties of the thinned Al films.
	Critical in-plane magnetic field, $B_{c,||}$, for sample 1 is shown in Fig.~\ref{fig3}(a) as a function of temperature, $T$. The GaAs substrate of sample 1 is insulating at low temperature. Four-terminal longitudinal resistivity, $\rho_{xx}$, was measured on two devices, $AO_I$, $AO_{II}$, both anodized at 4 V, and on a third device with Al as grown $(AG)$ without AO.
	For the two $AO$ samples we observe $B_{c,||} > 6$~T at base temperature and $B_{c,||}=6$~T at $1.2$~K, while the $AG$ shows $B_{c,||} = 3$~T at base temperature.  Figure~\ref{fig3} indicates that the two AO devices were essentially identical. Additional scans are displayed in SM Sec.~E \cite{SM}. 
	For the $AG$ device, the normal state resistivity is low, $\rho_{N,AG}=3.5\text{ }\Omega/\square$, and the transition is sharp, while the $AO$ Hall bars have higher resistivity, $\rho_{N,AO}\sim3.5\text{ k}\Omega/\square$, and a broad transition. The thousandfold increase in normal-state resistivity in the $AO$ film compared to the $AG$ film presumably reflects increased scattering \cite{Wissmann2007} and inhomogeneity, not simply reduced thickness. 
	$B_{c,||}$ was defined as $\rho_{xx}(B_{c,||})=0.01\, \rho_{N}$.
Temperature sweeps were also used to extract $T_c(B_\parallel=0)$. $AG$ has $T_c=1.6\text{ K}$ while $AO_I$ has $T_c=2.3\text{ K}$.
	
	For a BSC-like superconductor with $T_c=2.3\text{ K}$, the theoretical Chandrasekhar-Clogston (CC) limit \cite{Chandrasekhar1962,Clogston1962} for the critical field, assuming a metallic Lande g-factor of 2, is $B_c(T=0) \sim 4.6$~T, which is clearly violated in these samples. Violation of CC limit has earlier been reported on a Pb film \cite{Nam2016}, where the authors claim that the spins are momentum locked in Rashba subbands due to high spin-orbit energy, which suppresses Zeeman splitting.  A related suppression of Zeeman splitting could originate from the high surface-to-bulk ratio of the anodized films, which could lead to large Rashba spin-orbit coupling within the film  \cite{Dimitrova2007}.
	
	Critical perpendicular magnetic field as a function of temperature $B_{c,\perp}(T)$, is displayed in Fig.~\ref{fig3}(b-c). We determine $B_{c,\perp}$ as the field for which $\rho_{xx}(B_{c,\perp})=0.01 \, \rho_{N}$. 
	A base-temperature scan of the $AG$ sample yielded $B_{c,\perp}=32 \text{ mT}$, higher than the bulk Al value $B_c\sim 10\text{ mT}$ \cite{Ihn2005}. 
	Scans used to determine $B_{c,\perp}(T)$ for $AO_I$ are shown in Fig.~\ref{fig3}(b). Extracted critical fields for $AO_I$ and $AO_{II}$ are shown in Fig.~\ref{fig3}(c) ($AO_{II}$ scans are shown in SM Sec.~E \cite{SM}). A nearly two orders of magnitude increase in $B_{c,\perp}=2.5$ T is observed. This is comparable with $B_{c,\perp}$ extrapolated from measurements on granular aluminum \cite{Cohen1968,Chui1981}. 
	
	Just below 3~T, a crossover point for different isotherms is observed. Such crossovers indicate superconductor-insulator transitions, studied in thin and/or disordered superconductors 
	\cite{Ando1995,Gantmakher2000,Makise2009,Xing2015,Boettcher2018}. A similar crossing was not observed in the probed range of $B_\parallel(T)$, but it might occur beyond 6 T.
	The resistance jumps seen below $1\text{ T}$ at high temperatures track temperature fluctuations, which were monitored simultaneously.\\
	Comparing $B_{c,||}(T)$ and $B_{c,\perp}(T)$, the latter has a more linear $T$-dependence. This is expected from Ginzburg-Landau (GL) theory and previously observed in thin quench-condensed Al films \cite{Grassie1971}.
	Another feature worth highlighting is the upward curvature of $B_{c,\perp}$(T) for $T\lesssim T_c$, previously observed in thin granular aluminum \cite{Shinozaki1983}, where the authors attributed the effect to electron localization and electron-electron interaction. \\

	\section{Anodized A\lowercase{l} on I\lowercase{n}A\lowercase{s} 2DEG: \\ Transport Studies}
	\label{InAs}
	
	\begin{figure}
		\includegraphics[width=\linewidth]{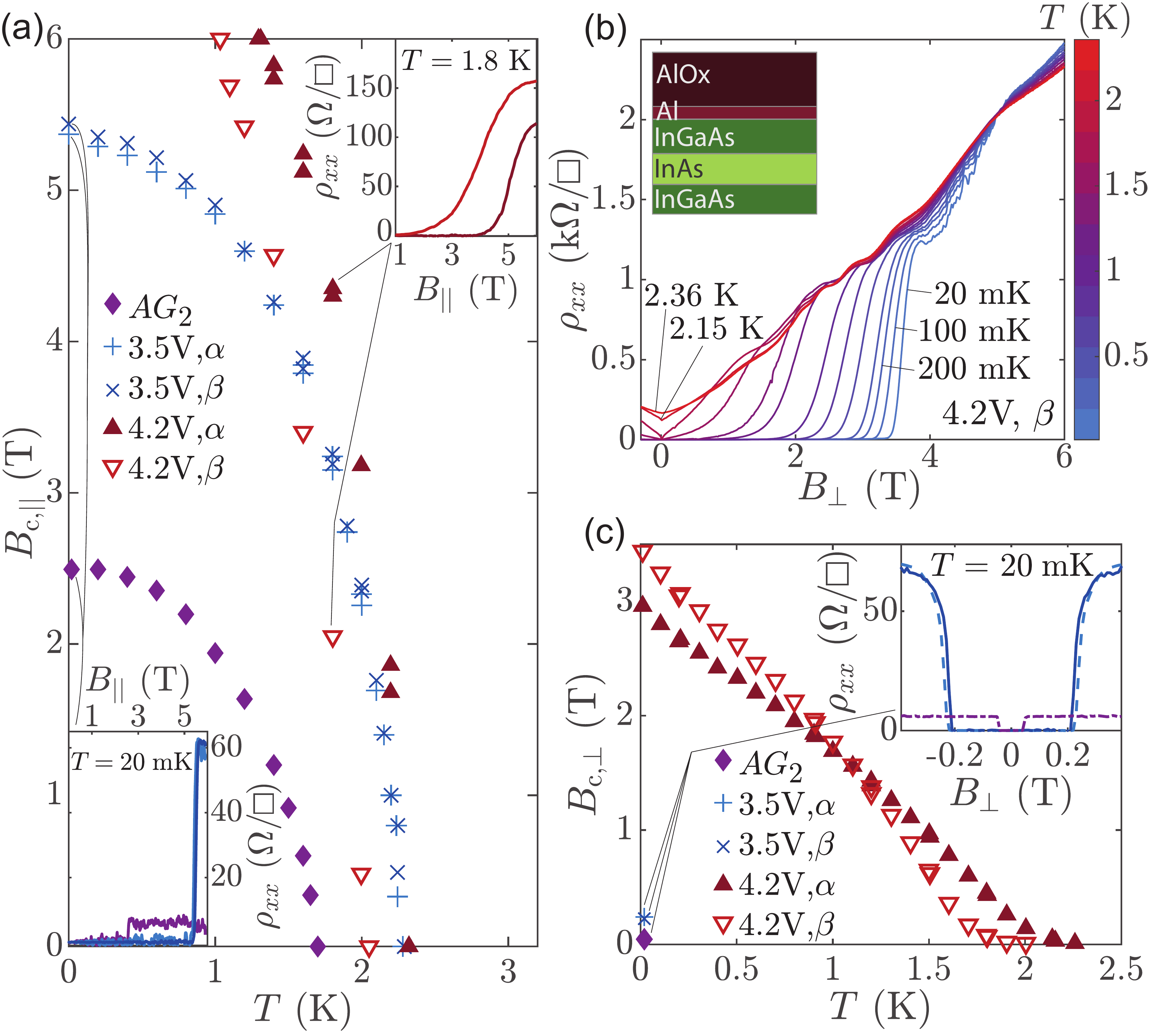}
		
		\caption{Transport data from sample 2.		
			\textbf{(a)} Critical in-plane field, $B_{c,||}$, as a function of temperature, $T$, for as grown Al, $AG_2$, and $V_{\rm AO} = 3.5$ and 4.2~V Hall bars on devices $\alpha$ and $\beta$.
			Some of the 3.5 and 4.2 V points were measured twice.
			Examples of $\rho_{xx}(B_\parallel)$ scans are included in the insets, taken at the stated temperatures.
			\textbf{(b)} $\rho_{xx}$ as a function of $B_\perp$ at different temperatures for the anodized Al film at 4.2 V. Inset: Heterostructure of sample 2.
			\textbf{(c)} Extraction of $B_{c,\perp}(T)$ for both Al films anodized at 4.2 V including base temperature $B_{c,\perp}$ for both Hall bars anodized at 3.5 V and for $AG_2$, extracted from scans displayed in the inset. 
		}
		
		\label{fig4}
		
	\end{figure}

	Transport measurements of sample 2, comprising AO of epitaxial Al on a heterostructure with a shallow InAs 2DEG, show similar trends in field and temperature as Al on insulating GaAs, with the addition of transport signatures from the 2DEG at $B_\perp>B_{c,\perp}$. Figure~\ref{fig4}(a) shows $B_{c,||}(T)$ for Hall bars from device $\alpha$ and $\beta$, anodized at $V_\mathrm{AO}=$ 3.5 V and 4.2 V respectively, as well as a Hall bar with as-grown aluminum, $(AG_2)$, the subscript indicating sample 2. 
	The data is extracted from $\rho_{xx}(B_\parallel)$ scans at fixed temperature. Lower-left inset displays $\rho_{xx}(B_\parallel)$ from $AG_2$ and 3.5 V Hall bars measured at 20 mK, showing a sharp superconducting transition. Top-right inset presents 4.2 V measurements of $\rho_{xx}(B_\parallel)$ at 1.8 K, showing broad superconducting transitions. The full set of measurements are included in SM Sec.~F \cite{SM}.
	
	$AG_2$ has a normal state resistivity $\rho_{N,AG}=6\,\Omega/\square$. On both $\alpha$ and $\beta$, the Al anodized with 3.5 V has $\rho_{N,3.5,\alpha}=\rho_{N,3.5,\beta}=61\,\Omega/\square$. The Al anodized with 4.2 V has $\rho_{N,4.2,\alpha}=120\,\Omega/\square$ and $\rho_{N,4.2,\beta}=160\,\Omega/\square$ and broad superconducting transitions. Here, the critical fields are again extracted as $\rho_{xx}(B_{||,c})=0.01\,\rho_{N}$. $T_c$ at 0 and 6 T was obtained through temperature sweeps at those fields.
	The critical field and temperature for $AG_2$ are $B_{c,||}=2.5 \text{ T}$ and $T_{c}=1.7$ K while for the 3.5 V films, $B_{c,||}=5.4$~T and $T_{c}=2.3$ K. 
	The thinnest films, using 4.2~V AO on $\alpha$ and $\beta$, have in-plane critical fields exceeding 6~T. 
	The critical temperature at 0 and 6 T are 2.3 K and 1.3 K for 4.2 V on device $\alpha$ and 2.1 K and 1.0 K for 4.2 V on device $\beta$.

	Both of the sample-2 AO Hall bars using $V_\mathrm{AO} = 4.2$~V have resistivities an order of magnitude lower than the resistivity of $AO_I$ on GaAs, due to parallel conduction through the 2DEG.
	The fact that $\rho_{N,4.2,\beta}>\rho_{N,4.2,\alpha}$ indicate that the Hall bar $\beta$ has a thinner or more disordered Al film. It also has a lower $T_c$ indicating that we are beyond the optimal thickness/disorder for $T_c$ after which it is expected to drop \cite{Ivry2014,Cohen1968}. 
	The notable difference between the two 4.2 V Hall bars indicates a lack of repeatability. For these few-ML thick aluminum films a small change in thickness/disorder have a large impact on the transport, making these films good probes for repeatability. The devices were made simultaneously from the same fabrication steps. Discrepancies could arise from sensitivity to the time over which AO was carried out, or due initial non-uniformity in the Al thickness, even in the ML range.

	Figure~\ref{fig4}(b) shows $\rho_{xx}(B_\perp)$ scans at different temperatures of the $V_\mathrm{AO}=4.2\text{ V, }\beta$ device, while data from $V_\mathrm{AO}=4.2\text{ V, }\alpha$ is shown in SM Sec.~F \cite{SM}. In both cases we again see a crossing of the isotherms, but on the GaAs substrate, Fig.~\ref{fig3}(b), $\rho_{xx}(B_\perp)$ converges shortly after the crossing. On InAs, at least up to 6 T, $\rho_{xx}(B_\perp)$ keeps changing, possibly due to $B_\perp$-induced orbital effects in the 2DEG, which we investigate in \cite{Drachmann2020}.
	
	Even though the normal state resistances increase in $B_\perp$, to extract $B_{c,\perp}(T)$, we use $1\%$ of the same $\rho_{N}$ that was used when extracting $B_{c,||}(T)$. 
	The result is displayed on Fig.~\ref{fig4}(c) together with extractions from the other four Hall bars. Both of the $3.5\text{ V}$ and the $AG_2$ Hall bar were only measured at base temperature, line scans shown in the inset.
	$B_{c,\perp}(T\sim20\text{ mK})=40$ mT for $AG_2$ Hall bar, $230$ mT for the $3.5\text{ V}$ Hall bars, while $4.2 \text{ V, }\alpha$ and $\beta$ has $B_{c,\perp}=3.0$ T and $3.5$ T, respectively.
	As observed for anodized Al on GaAs substrate, Fig.~\ref{fig3}(c), $B_{c,\perp}$ is increased by more than two orders of magnitude and for the thinnest Al Hall bars, the $B_{c,\perp}(T)$ dependence is close to linear with an upward curvature close to $T_c$.


	\section{Passivation of shallow I\lowercase{n}A\lowercase{s} 2DEG}

	\begin{figure}
		\includegraphics[width=\linewidth]{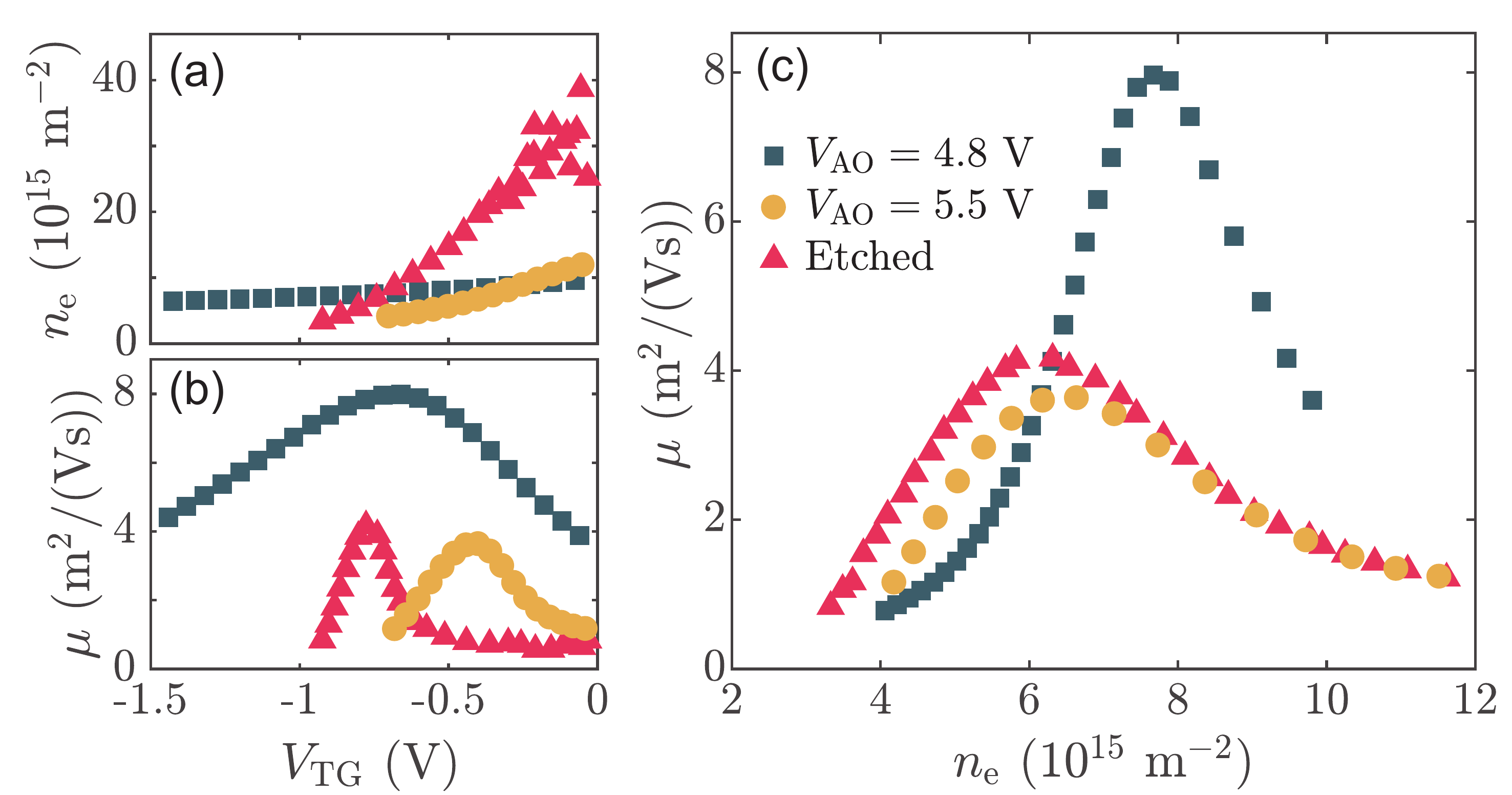}
		
		\caption{
			Dependence of mobility, $\mu$, and electron density, $n_{\rm e}$, on top-gate voltage, $V_\mathrm{TG}$, for three devices from sample 2, identified by legend in (c). The red triangle is from a device where Al was chemically wet etched. The  yellow circles and blue squares are from Hall bars anodized at 5.5~V and 4.8~V respectively.
			\textbf{(a)}  Density as a function of top-gate voltage.
			\textbf{(b)} Mobility as a function of top-gate voltage.
			\textbf{(c)} Parametric plot of mobility as a function of density.	
		}
		
		\label{fig5}
		
	\end{figure}

	\label{IIIV}

	Further increase of $V_\mathrm{AO}$ resulted in Hall bars having finite resistance at base temperature, indicating a full oxidation of the Al film. Such an outcome allows the semiconductor underneath to be passivated and 2DEG density to be controlled with the top-gate.  Top gating was found to be possible on Hall bars anodized at $V=4.8$~V and 5.5~V, see Fig.~\ref{fig5}(a-b), indicating that the residual metallic Al film was either discontinuous or completely oxidized.

	\begin{figure*}
		\includegraphics[width=0.82\linewidth]{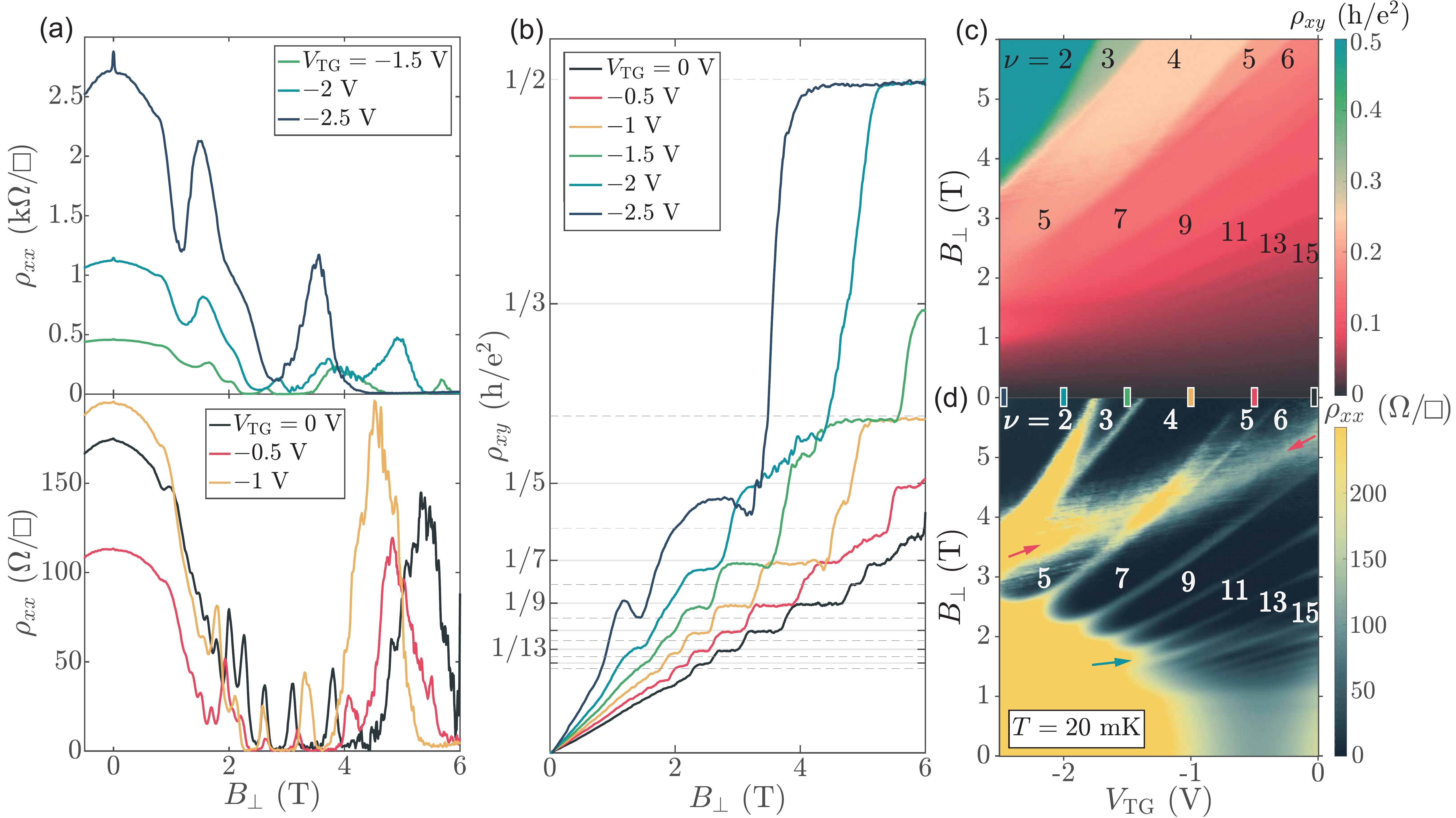}
		
		\caption{
			Quantum Hall regime at base temperature for Hall bar anodized at 4.8 V.
			\textbf{(a)} Longitudinal and \textbf{(b)} Hall resistivities as a function of $B_\perp$ for different top-gate voltages. 
			In (b) the quantum Hall resistivities for odd (solid) and even (dashed) integer filling factors marked. 
			2D map of the \textbf{(c)} Hall  and \textbf{(d)} longitudinal  resistivities as a function of $B_\perp$ and top-gate voltage, $V_\mathrm{TG}$. Extracted filling factors are indicated on the maps. 
			Arrows in (d) highlight additional resonances.
		}
		
		\label{fig6}
		
	\end{figure*}

	Characterization of the two Hall bars was done measuring longitudinal resistivity $\rho_{xx}$ and Hall resistivity $\rho_{xy}=dV_{xy}/dI$ as a function of $B_\perp$ and top-gate voltage, $V_\mathrm{TG}$, to extract carrier density, $n_e$, and carrier mobility, $\mu$, from the low field Hall effect. The same analysis has been performed on a Hall bar fabricated with standard chemical wet etch of the Al film. 
	The $V_\mathrm{TG}$ dependence of mobility, $\mu$, versus density, $n_\mathrm{e}$ for the three Hall bars is shown in Fig.~\ref{fig5}.
	Details concerning the fabrication of the etched Hall bar and additional quantum Hall data of another Hall bar, also anodized at 4.8 V, are given in SM Sec.~G \cite{SM}.

	Carrier density as a function of $V_\mathrm{TG}$ is shown in Fig.~\ref{fig5}(a). For all Hall bars, $n_\mathrm{e}$ decreased monotonically with negative gate voltage $V_\mathrm{TG}$ (the etched Hall bar was noisy for $V_{TG}>-0.3$ V). 
	At $V_\mathrm{TG}=0$, the etched Hall bar has $n_\mathrm{e}\sim4\times10^{12}$~$\mathrm{ cm^{-2}}$, while the Hall bars anodized at 5.5 V and 4.8 V has $n_\mathrm{e}=1.1\times10^{12}$~$\mathrm{ cm^{-2}}$ and $n_\mathrm{e}=9.8\times10^{11}$~$\mathrm{ cm^{-2}}$, respectively. The same order is observed in the slope of $n_\mathrm{e}$ versus $V_\mathrm{TG}$, with the etched Hall bar having the steepest slope and the 4.8 V Hall bar slope being most flat.
	Corresponding mobilities are shown in Fig.~\ref{fig5}(b) as a function of $V_\mathrm{TG}$ and in Fig.~\ref{fig5}(c) as function of $n_\mathrm{e}$.
	We speculate on the origin of the nonmonotonic dependence of mobility with density: 
	At low densities, the initial increase in $\mu$ with increasing $n_\mathrm{e}$ presumably reflects an increased range of validity for Thomas-Fermi screening.
	A peak in mobility followed by a decrease with further increase of $n_\mathrm{e}$ presumably reflects a greater concentration of electrons at the surface.

	The wet-etched Hall bar has a mobility peak of $4.2\time10^4$~$\text{cm}^2/\text{Vs}$ at $n_\mathrm{e}=6.1\times10^{11}$~$\mathrm{cm^{-2}}$, which is high for a shallow 2DEG structure, taking into account that this 2DEG has transparent coupling to the Al, as shown in Ref. \cite{Fornieri2019}. The Hall bar anodized at 4.8 V has an increased mobility peak of $8.0\times10^4$~$\text{cm}^2/\text{Vs}$ at $n_\mathrm{e}=7.7\times10^{11}$~$\mathrm{cm^{-2}}$ while the $5.5$ V AO Hall bar has a reduced mobility peak of $3.7\times10^4~\text{cm}^2/\text{Vs}$ at $n_\mathrm{e}=6.5\times10^{11}$~$\mathrm{cm^{-2}}$. 
	It is known that the aluminum etch damages the underlying semiconductor \cite{Shabani2016a}, partially due to the exposure to atmosphere \cite{Pauka2019}. 
	We attribute the increased peak mobility after $4.8$ V AO to an alumina passivation of the III/V, as depicted in Fig.~\ref{fig1}(g), decreasing surface scattering compared to the etched Hall bar. 
	AO at $5.5$ V showed decreased peak mobility compared to the etched Hall bar.  We suspect that such a voltage introduces disorder by contributing to oxidation of the semiconductor underneath, as depicted in Fig.~\ref{fig1}(g).

	Measurements of $\rho_{xx}$ and $\rho_{xy}$ as a function of $V_\mathrm{TG}$ and $B_\perp$ up to 6 T are shown in Fig.~\ref{fig6}. These data are for the highest mobility Hall bar, anodized at 4.8 V.
	Fig.~\ref{fig6}(a) shows $\rho_{xx}$ as a function of $B_\perp$ for different $V_\mathrm{TG}$ between 0 and $-2.5$~V. 
	Shubnikov de Haas oscillations (SdHO) are visible for $B_\perp > 1$ T with vanishing minima of $\rho_{xx}$ for $B_\perp \sim 2.5$ T. We emphasize that this value of field is $\sim$ 1~T {\it lower} than our highest superconducting $B_{c,\perp}$.
	Along with vanishing $\rho_{xx}$, quantized plateaus of $\rho_{xy}$ are observed in Figure 6(b) for $\rho_{xy}$ with $B_\perp>2$ T, indicating that the sample is within the quantum Hall regime at fields below $B_{c,\perp}$.

	Besides SdHO, extra $\rho_{xx}$ peaks are present in Fig.~\ref{fig6}(a). Starting at the least negative $V_{TG}$, a broad and tall resistivity spike is observed between 4 and 6 T and a smaller increment to $\rho_{xx}$ is observed just below 2 T. For the most negative $V_{TG}$, the high field spike moves towards 3 T, and the increment below 2 T becomes more prominent. Lastly a weak localization peak \cite{Altshuler1980} is observed around $B_\perp=0$.

	Additional features are observed in the Hall resistivity as well. At fields where $\rho_{xx}$ has additional spikes, $\rho_{xy}$ is noisy and forms non-integer plateaus. Below $4$~T, $\rho_{xy}$ only has plateaus at odd fractions. This is seen more clearly in Fig.~\ref{fig6}(c), displaying a color map of $\rho_{xy}(V_\mathrm{TG},B_\perp)$, from which we extract the filling factor $\nu$ and display it across the map. The same $\nu$ are added to a color map of $\rho_{xx}(V_\mathrm{TG},B_\perp)$, Fig.~\ref{fig6}(d), showing a Landau fan diagram with additional features. The most obvious of these, a resonance crossing the fan at high fields, marked by red arrows, is the origin of the high field resistance spikes mentioned above. We notice that above the resonance, $\nu$ has regular even integer counting with spin-splitting into odd values around $B_\perp\sim5$~T. Below the resonance, $\nu$ obtains odd integer counting. Another, less prominent, resonance, marked by the blue arrow, is the origin of the increment and spikes below 2 T in Fig.~\ref{fig6}(a). A better visualization of the low-field features is obtained from an un-saturated color map of $\rho_{xx}(V_\mathrm{TG},B_\perp)$ included in SM Sec.~H \cite{SM}.

	A similar high field study, presensted in SM Sec.~H \cite{SM}, was done on the other Hall bar anodized at 4.8 V which had a mobility of $6.7\times10^4~\mathrm{cm^2/Vs}$ at a density of $n_e=7.6\times10^{11}$ $\mathrm{cm^{-2}}$.
	Here, as well, we observe similar $\rho_{xx}$ resonances and odd $\rho_{xy}$ counting turning into even counting when crossing a high field resonance.

	Speculations about the rise of extra resonances in $\rho_{xx}$ and the addition to $\nu$ that comes with them, lead to an interpretation including a low density and disordered second band.
	The extra resonances represent its fan-diagram. The broad fans indicate low mobility.
	To explain the observed combined filling factor $\nu$, we suggest that crossing the high field resonance, marked by red arrows in Fig.~\ref{fig6}(d), takes the second band filling factor from $\nu_{2nd}=1$ to 0. Depletion of a Hall bar with $B_\perp$ is commonly observed in the study of disordered 2DEGs \cite{Glozman1995}. The origin of such a second band and where it would reside is not clear. It could be related to an unintentional quantum well deeper in the heterostructure.
	


	\begin{figure*}
		\includegraphics[width=\linewidth]{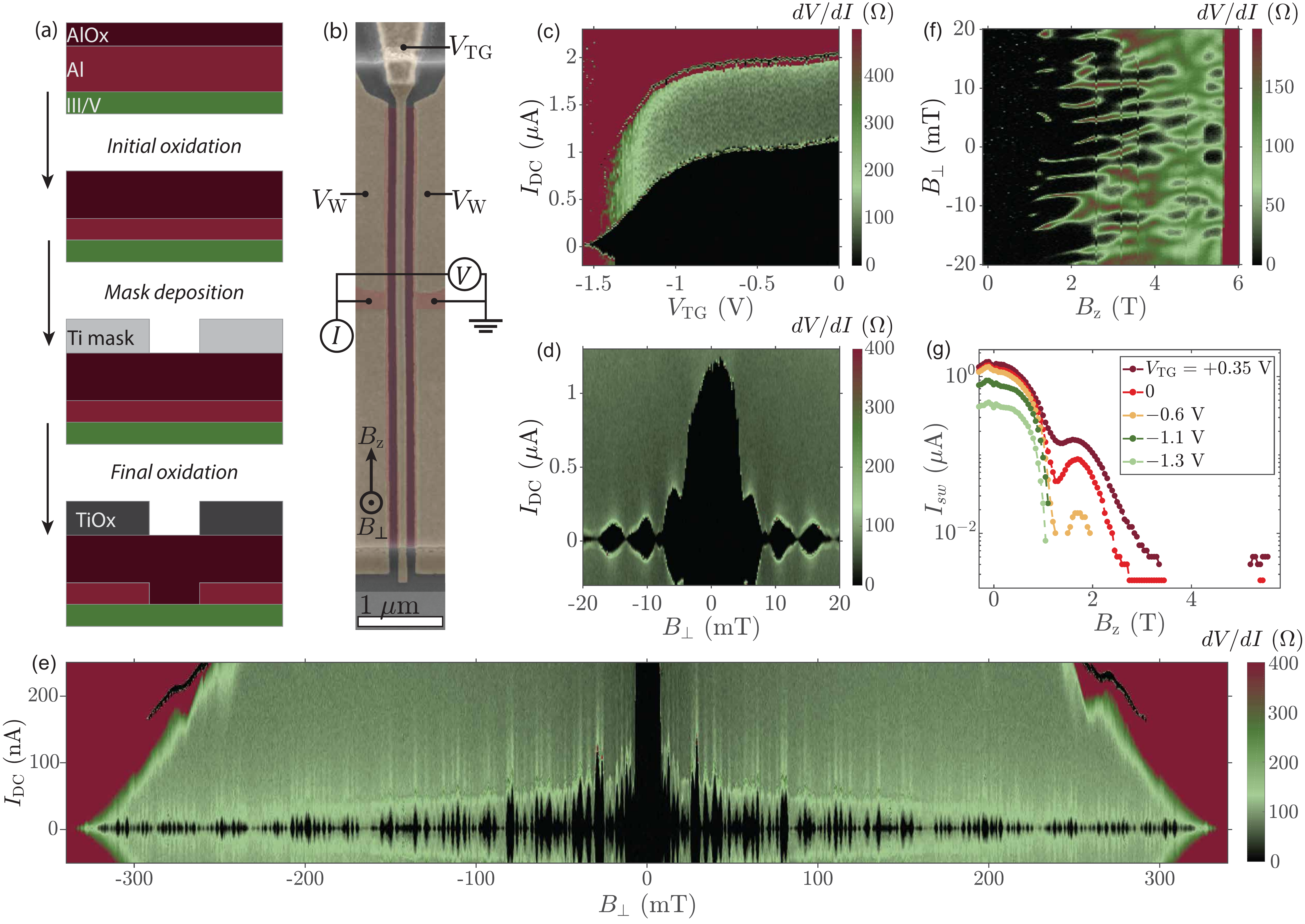}
		
		\caption{
			\textbf{(a)} Schematic illustration of the fabrication process combining standard lithography techniques with metal masks and anodic oxidation. 
			\textbf{(b)} False-color scanning electron micrograph of the reported narrow Josephson junction. 
			The remaining thin Al layer is in red, and the gates are represented in yellow. The two side gates $V_\mathrm{W}$ are kept at $-1.6$~V for the experiments.
			Transport was current-biased with voltage drop, $V$, measured in a four-terminal configuration. Field directions $B_\perp$ and $B_z$ are defined in the bottom left part of the figure.
			\textbf{(c)} Saturated color map of the differential resistance $dV/dI$ as a function of current bias $I_\mathrm{DC}$ and top-gate voltage $V_\mathrm{TG}$. 
			\textbf{(d)} Saturated color map of $dV/dI$ as a function of current bias and perpendicular field for $V_\mathrm{TG}=0$. 
			\textbf{(e)} Extended color map of $dI/dV$ as a function of $I_\mathrm{DC}$ and $B_\perp$ showing supercurrent oscillations surviving high fields $B_\perp>300$ mT, with $V_\mathrm{TG}=350$ mV.
			\textbf{(f)} Saturated color map of $dV/dI$ at zero bias as a function of $B_\perp$ and $B_z$, showing that the super-current is oscillating as a function of both field directions, with $V_\mathrm{TG}=0$. 
			\textbf{(g)} Oscillation and re-emergence of switching current, $I_\mathrm{sw}$ (see main text) vs $B_{z}$ for several top-gate voltages, $V_\mathrm{TG}$.
		}
		
		\label{fig7}
		
	\end{figure*}

	\section{Patterning Anodic Oxidation}
	\label{SNS}
	
	Combining lithography with anodic oxidation opens a new path toward patterning hybrid heterostructures. 
	In this section, we demonstrated patterning of the AO process, implemented on the InAs 2DEG heterostructure, sample 2. 
	Following a few unsuccessful approaches to patterning, described in SM Sec.~I \cite{SM}, a successful metal-mask approach was implemented, as shown in Fig.~\ref{fig7}(a). Following an initial global AO step, a thin Ti metal mask was patterned using electron beam lithography and liftoff, with Ti in the locations where Al should stay metallic. A second AO fully oxidizes the Al not covered by the mask as well as oxidizing the mask itself. Resulting structures are shown in SM Sec.~J \cite{SM}.
	The initial oxidation was done at $3.5$ V for 5 min. The resulting Al film had $B_{c,||}\approx5.6$ T and $B_{c,\perp}\approx300$~mT, similar to the $3.5$ V exposures of Fig.~\ref{fig4}, indicating a similar thickness.
	A final oxidation  was done at $4.8$ V for 5 min. The Hall bar showed a mobility peak of $1.16~\mathrm{m^2/Vs}$ at a density of $7.2\times10^{11}$ $\mathrm{ cm^{-2}}$. The mobility is much lower than the one reported for Hall bars in Fig.~\ref{fig5}, suggesting the oxidation process was too deep and impacted the semiconductor. 
	An unintended oxidation depth could be due to a time-dependent AO process, see SM Sec.~A \cite{SM}.
	The final oxidation, having the largest $V_\mathrm{AO}$, was intended to solely control the oxidation thickness. With a time-dependence, the depth of the initial and final oxidation would add up, making our current setup non-ideal for this purpose.

This method was used to fabricate 5~$\mu\mathrm{m}$ long superconductor-normal-superconductor Josephson junction (JJ) with 100 nm wide superconducting leads and normal regions varied between 100~nm and 400~nm, as shown in see Fig.~\ref{fig7}(b) and SM Sec.~K \cite{SM}.

	Measurement setup and applied field directions are also shown in Fig.~\ref{fig7}(b). An out-of-plane field $B_\perp$ is used in combination with an in-plane field $B_z$, perpendicular to the junction. An AC current bias of 0.5 nA was applied with possibility of adding DC bias, $I_\mathrm{DC}$, while an AC voltage drop was measured. A top-gate $V_{\mathrm{TG}}$ was used to deplete or populate the junction with carriers. The two wire gates $V_\mathrm{W}$, biased at $=-1.6$ V, were used to form the narrow junction in the InAs 2DEG enabled by electrostatic screening of the metallic aluminum \cite{Fornieri2019}.
	
	Gate control of the supercurrent is shown in Fig.~\ref{fig7}(c). Differential resistance ($dV/dI$) is measured as a function of $I_\mathrm{DC}$ and $V_\mathrm{TG}$, which related a critical current $I_c\sim1.1$ $\mu$A at $V_\mathrm{TG}=0$ V. 
	The Fraunhofer-like pattern of oscillations of critical current as function of perpendicular magnetic field,  $B_\perp$ is shown in Fig.~7(d). We observe that the critical current oscillates in $B_\perp$ as seen previously \cite{Suominen2017, Drachmann2017, Fornieri2019, Mayer2019}.
	For an ideal junction, the critical current is expected to follow:
	\begin{equation}
		I_c(B_\perp)=I_c(0)\left|\frac{\sin(\pi B_\perp A/\phi_0)}{\pi B_\perp A/\phi_0}\right|,
		\label{eq1}
	\end{equation}
	$\phi_0$ being the flux quantum and $A$ being the junction area which, in our case, is designed to be $\sim0.1$~$\mathrm{\mu m}\times5$~$\mathrm{\mu m}=0.50$~$\mathrm{\mu m}^2$.
	The observed period of $\sim4.5$ mT corresponds to an area of $0.46$~$\mathrm{\mu m}^2$, not far from the designed value.
	
	The perpendicular critical field is estimated to be 310~mT as evidenced in the large field Fraunhofer pattern of Fig.~\ref{fig7}(e). This value is consistent with the critical field of full-sheet anodized Al, see SM sec. J \cite{SM}. 	
	
	The narrow JJs were made to verify the resolution of the AO lithography with metal masks. A gateable junction verifies that the Al leads are not shorted and the Fraunhofer period indicates that the leads are not broken. This indicates that the horizontal lithographic resolution is better than 50 nm, both away from and in-under the metal mask.
	
Differential resistance $dV/dI$ was measured at $I_{DC}=0$ as a function of $B_z$ and $B_\perp$ with $V_\mathrm{TG}=0$, displayed on Fig.~\ref{fig7}(f). Similar maps were taken at other $V_\mathrm{TG}$ values, see SM Sec.~L \cite{SM}. The field directions were aligned to the device axes before data acquisition. A complicated pattern is observed with quenching and reappearance of the supercurrent as a function of both field directions. 
	For $B_z<1$ T, supercurrent persists for all $B_\perp\in[-20:20]$ mT. Beyond 1 T resistive regimes emerges and broadens out. No supercurrent is present after $B_z>4$ T, besides two superconducting blobs appearing at high fields $B_z\sim5.5$ T, just before the Al goes normal.
	
	Switching current, $I_{sw}$, defined as the maximum critical current for any $B_\perp$ in a 10~mT interval, as a function of $B_{z}$ between $-0.3$ T and $5.8$ T for  $V_\mathrm{TG}$ between $+0.3$~V and $-1.3$~V is shown in Fig.~\ref{fig7}(g).
	For $V_\mathrm{TG}=-1.3$ V and $-1.1$~V, we observe the supercurrent being quenched around $B_z\sim1$ T. For $V_\mathrm{TG}=-0.6$ V, the junction goes resistive as well around $B_z\sim1$ T, but superconductivity reemerges between $1.5$ and $2$ T. For $V_\mathrm{TG}=0$ and $+0.35$ V, $I_{sw}$ sustains up to $\sim3.4$ T, with a local peak around $1.7$ T. A finite $I_{sw}$ was also observed for $B_{z}$ between $5.2$ and $5.5$ T, close to the critical field of the superconductor itself. Resistance of the junction in the normal regime also oscillates with $B_z$ but out of phase with the $I_{sw}$ oscillations, as can be seen in SM Sec.~L \cite{SM}.
	We observe that the $B_z$ values for peaks and dips in $I_{sw}$ are independent of $V_{TG}$ and thus also the InAs chemical potential. The $B_z$ value resulting in a topological phase transition is expected to have a chemical potential dependence \cite{Hell2017, Pientka2017}. We therefor attribute the oscillations to be $B_z$-induced orbital effects as reported in \cite{Fornieri2019}. For an orbital effect, periodic nodes and anti-nodes should be expected. With $I_{sw}$ peaks at $B_z\sim1.7$ T and $\sim5.4$ T, we could be observing 1st and 3rd anti-node, while a second anti-node around $B_z\sim3.5$~T could be suppressed.


	\section{Conclusion}
	
	In summary, we have developed an anodic oxidation approach to controlling the thickness of epitaxially grown Al on hybrid Al/InAs heterostructures. We apply this process to (i) thin epitaxially grown Al resulting in increased critical temperature and magnetic fields, (ii) passivate hybrid heterostructures with complete oxidation of the Al films so that the interface between the heterostructure and oxide has never been exposed to air, and (iii) pattern superconductor-normal junctions when combined with lithographic processes to produce nanoscale structured superconducting elements, focussing on the case of SNS junctions with sub-100 nm features.
	The anodized Al films showed critical fields $B_{c,||}>6 \text{ T}$ and $B_{c,\perp}\sim3.5 \text{ T}$. Resistivity of thin disordered Al yielded $\rho_{xx}>1\,\mathrm{k\Omega}$, making the material interesting for high-kinetic impedance superconducting resonators \cite{Maleeva2018}.
	Full oxidation of the Al increases peak mobility to $8\cdot10^4\mathrm{cm^2/Vs}$, twice the value obtained with regular Al etch. 
	Quantum Hall effect was observed at $\sim2.5$ T, giving $\sim$ 1 T overlap with highest superconducting critical field, enabling future research into superconductor-quantum Hall hybrids.	%
	Applying a metal mask, lithographic resolution $<50$~nm was achieved.
	The different possibilities offered by the anodic oxidation process combined with the improved properties of the resulting films motivate further research in anodic oxidation or other new fabrication processes for hybrid heterostructures. This opens the path towards new applications and new regimes for experiments on hybrid heterostructures.

	\section{Acknowledgements}
	The authors acknowledge Ahbishek Banerjee, Mikhail Feigel'man, Joshua Folk, Roman Lutchen and Christoph Strunk for valuable discussions, and Saeed Fallahi, Tailung Wu and Teng Zhang for assistance with measurements.
	This work was supported by Microsoft Corporation, the Danish National Research Foundation, and the Villum Foundation.
	Fabrizio Nichele acknowledges support from European Research Commission, grant number 804273.
	STEM studies and all related sample preparation were done at the Birck Nanotechnology Center, Purdue University.

	\bibliography{AOpaper1}

\newpage

\newcommand{\beginsupplement}{%

\setcounter{figure}{0}
\renewcommand{\thefigure}{S\arabic{figure}}%

\setcounter{section}{0}
\renewcommand{\thesection}{\Alph{section}}:

}

\newpage

\onecolumngrid

\begin{center}
{\Large 	Supplementary Material}
\end{center}
\maketitle

\beginsupplement

\section{Anodic oxidation: Chemistry \& Electrical Setup}
\label{A}

The electrolyte we used for anodic oxidation (AO) in this work was tartaric acid, pH regulated på ammonium hydroxide, inspired by Ref.~[40]. We mixed a 3\% (w/w) tartaric acid powder and MQ water solution with a magnet stirrer at room temperature until all powder is dissolved. According to Ref.~[38], the acid should be pH buffered to a pH of 5-5.5 to obtain a barrier-type (uniform dielectric) oxide growth. This was done by carefully adding $\mathrm{NH_4OH }(10\%<\mathrm{konc.}<25\%)$, tracking the pH after each addition ending up with a pH of 5.45. 

\begin{figure}[b]
	\includegraphics[width=0.65\linewidth]{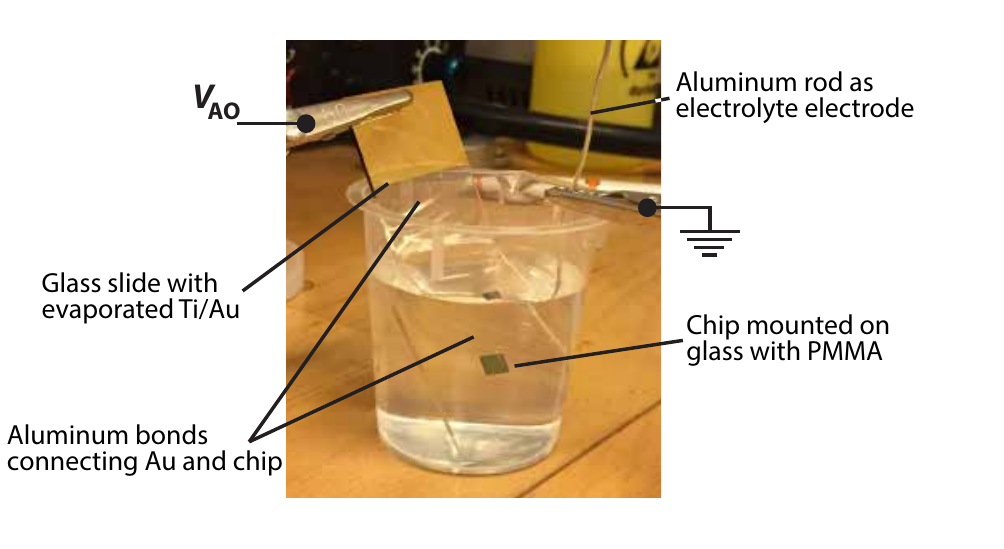}
	
	\caption{
		Electrical setup used for executing anodic oxidation. The chip was (with PMMA) mounted on a glass slide that have Ti/Au evaporated on the other end. Aluminum bonds are shorting the Ti/Au and chip. Glass slide and an aluminum rod are put in the electrolyte and are both contacted with alligator clips to source a potential difference $V_\mathrm{AO}$.
	}
	
	\label{figS1}
	
\end{figure}

Our electrical setup is displayed in Fig. \ref{figS1}. Electrical contact to our small chips was established through small aluminum bonds. To stabilize the setup a glass slide was prepared with Ti/Au evaporated in one end (the top). The chips would be mounted, with PMMA, to the bottom end and long aluminum wires were bonded from chip to the Ti/Au, which then can be contacted by an alligator clip connected to a voltage source. The glass slide with a chip bonded to Ti/Au was put into the mixed electrolyte together with an aluminum rod, grounded by another alligator clip. 

With the glass slide and rod submerged, the power supply was turned on and ramped with 100 mV/s to the desired voltage, $V_\mathrm{AO}$. Once reached, the process ran for 5 minutes before voltage was ramped down, again with 100 mV/s. Once at zero, alligators were unhooked and the glass slide was rinsed in MQ. The chip was then stripped from the glass using acetone and isopropanol.

With this setup, the oxidation depth was found not only to depend on the applied voltage, but to have an unintended time-dependence as well. This was realized at the end of this study. It might have been accounted for by trying other electrolyte solutions.

To verify a time-dependent oxidation, a test was conducted on the GaAs substrate, sample 1. With electron beam lithography a confined region was defined on a newly-cleaved chip. Two-terminal resistance was measured at room temperature with a DMM, through aluminum bonds, both prior to and after AO. Before anodic oxidation, a low resistance $R\sim15$ $\Omega$ of the metallic Al was measured. AO was done as described above, but at a low voltage $V_\mathrm{AO}=3.0$ V and with the process running for 25 min. After AO, the resistance was measured, through the same bonds, to be $R>1$ $\mathrm{M}\Omega$, indicating a full oxidation of the aluminum. With enough time, a low $V_\mathrm{AO}$ (lower than any of the values used for experiments in this work), could fully oxidize the aluminum, thus indicating a time dependence.

\newpage

\section{Fabricating Devices on GaAs \& Optical Inspection}
\label{B}
\begin{figure}[h]
	\includegraphics[width=0.6\linewidth]{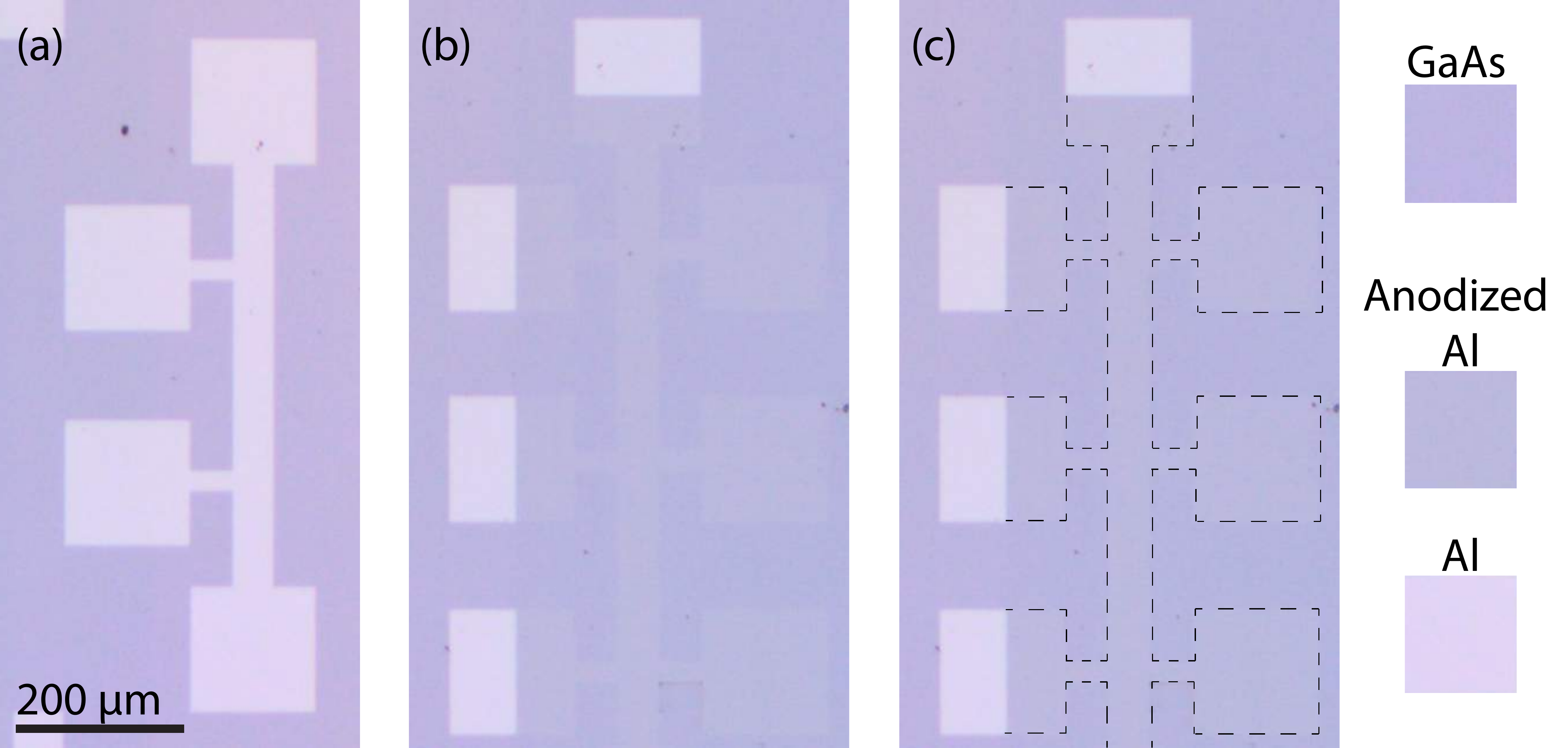}
	
	\caption{
		Optical micrographs showing devices measured on GaAs substrate. 
		\textbf{(a)} As grown $AG$ bar etched out of as-grown aluminum. Scale bar applies for all figures.
		\textbf{(b)} Anodic oxidation $AO$ Hall bar etched out in anodized region. Oxidized aluminum is hard to distinguish from the substrate so some Al was left unprocessed on bonding pads to enable localization of the pads. 
		\textbf{(c)} Identical to (b), but with dashed lines indicating boundary of Hall bar and substrate.
	}
	
	\label{figS2}
	
\end{figure}

On sample 1, epitaxial Al on GaAs substrate, the anodic oxidation was done, as described in Sec. \ref{A}, with 5~min at $V_\mathrm{AO}=4$~V in an electron beam lithography (EBL) defined area with a regular PMMA mask. Consecutively an EBL defined chemical wet etch (MF-321 developer for 5 min at room temperature) was done to define elongated Hall bars in the anodized and as-grown AG, see Fig. \ref{figS2}.

\section{Anodic oxidation \& fabrication on InAs 2DEG heterostructure}
\label{C}
\begin{figure}[h]
	\includegraphics[width=0.5\linewidth]{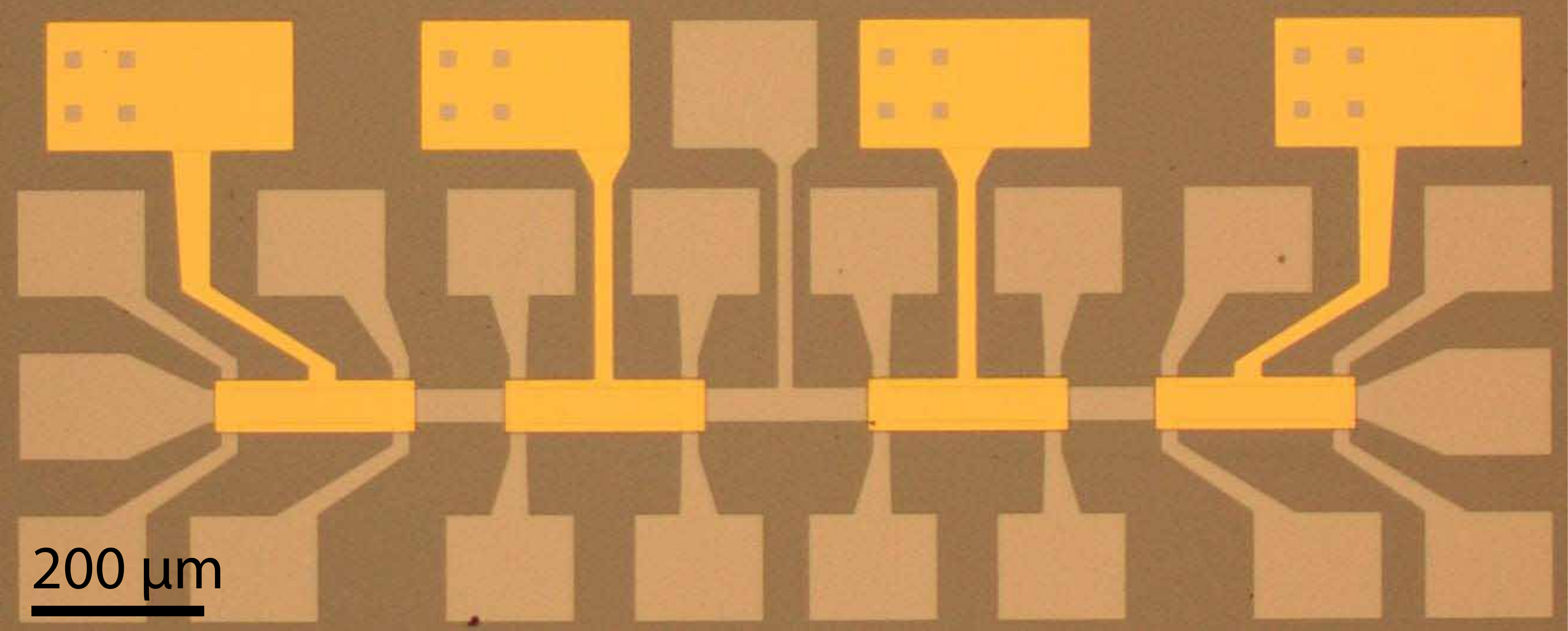}
	
	\caption{
		Optical micrograph showing same devices as Fig. 1(d) of main text, but here with the Ti/Au top-gates evaporated.
	}
	
	\label{figS3}
	
\end{figure}

Sample 2, Al grown epitaxially on a shallow InAs two dimensional electron gas (2DEG) heterostructure, was fabricated with the following steps:
\begin{itemize}
	\item Mesa definition
	\item 4 $\times$ AO, in four consecutive lithography steps
	\item Atomic layer deposition
	\item Top-gate evaporation
\end{itemize}
All lithography was done with standard electron beam techniques. 
Mesa etch was done by stripping the Al cap with 6~sec in Al etch Transene D, followed by a mesa etch solution:
$\mathrm{H_2O}$ : $\mathrm{C_6H_8O_7}$ : $\mathrm{H_3PO_4}$ : $\mathrm{H_2O_2}$ in a 220:55:3:3 ratio. 
After 9:30 min a depth of $\sim300$ nm was achieved.
Anodic oxidation was done as described in Sec. \ref{A}. Each AO was done in a separate EBL step, with the $V_{AO}=$ 3.5 V, 4.2 V, 4.8 V, and 5.5 V.
Hafnia was deposited at 90 degC, with 150 pulses of TDMAH and $\mathrm{H_2O}$ in a Cambrigde ALD system.
Lastly, patterned top gates of 10 nm Ti and 350 nm Au was e-beam evaporated on top of all devices, see Fig. \ref{figS3}.

\newpage

\section{STEM and EELS of AO on GaAs}
\label{D}
This section goes in depth with the scanning transmission electron microscopy (STEM) and electron energy loss spectroscopy (EELS) analysis done on Sample 1. This will be done in three subsections: (1) Characterizing growth direction of Al on GaAs, (2) STEM statistics and (3) EELS statistics.

\subsection{Growth direction of Al on GaAs}
\begin{figure*}[h]
	\includegraphics[width=0.7\linewidth]{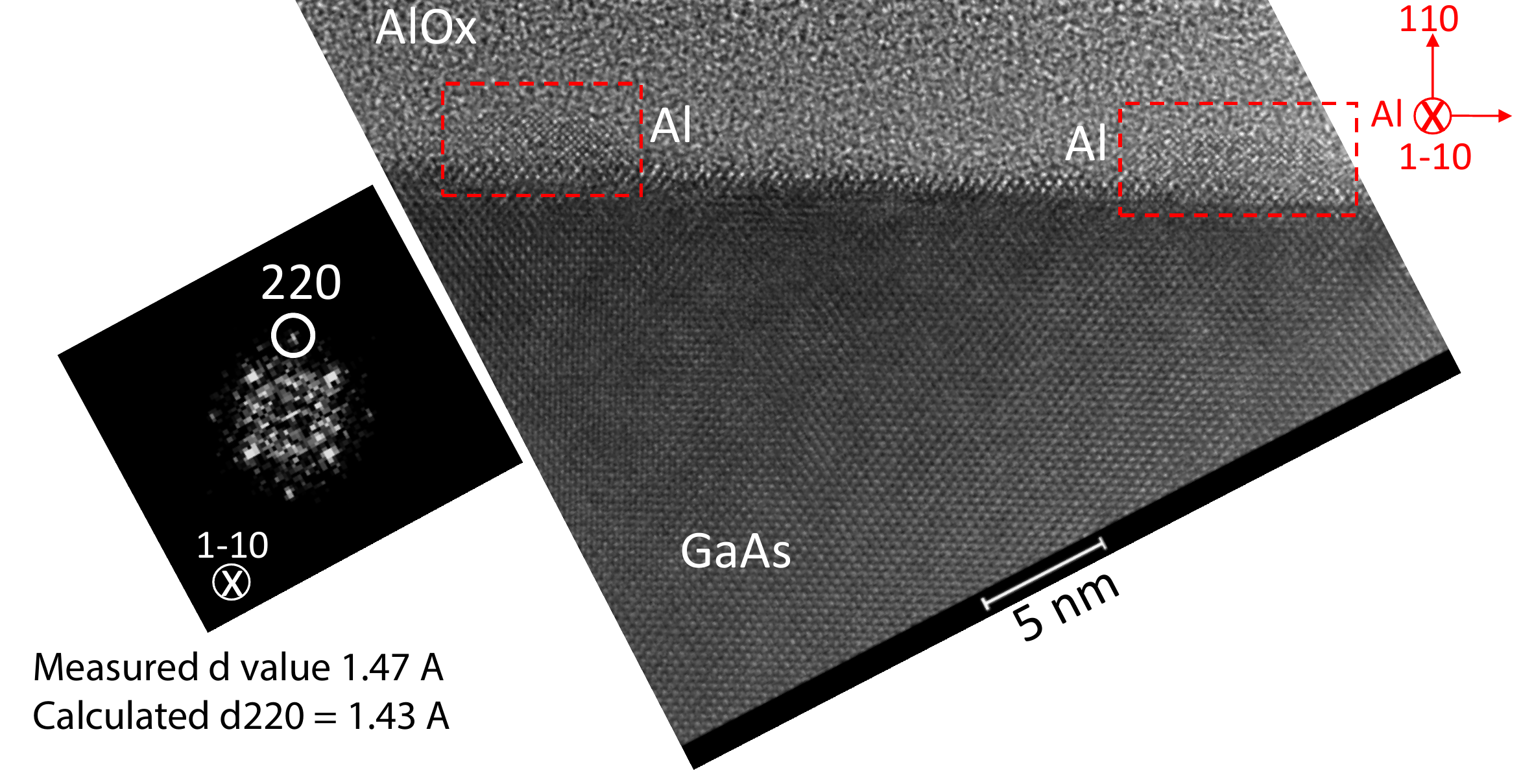}
	
	\caption{
		High-resolution transmission electron microscopy image showing the AlOx layer, Al hills at the GaAs interface, and the GaAs layer. Fast Fourier transform  of the Al hills shows that the Al crystal grows oriented in the [110] direction parallel to the [001] growth direction of the GaAs layer.
	}
	
	\label{figS4}
	
\end{figure*}

\newpage

\subsection{STEM statistics}
\begin{figure*}[h]
	\includegraphics[width=0.6\linewidth]{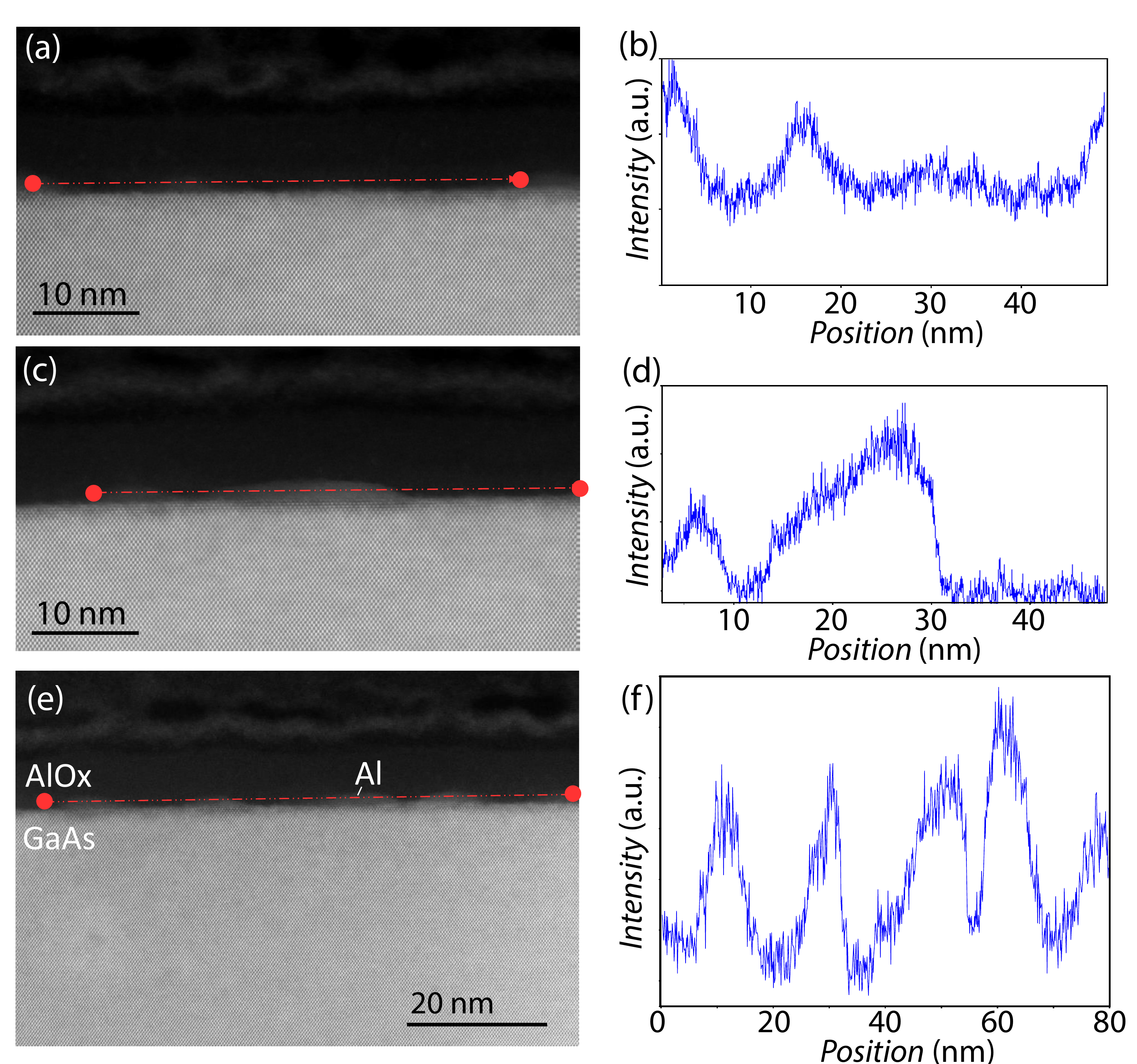}
	\caption{
		\textbf{(a,c,e)} High resolution scanning transmission electron microscopy images using a high-angle annular dark-field detector and a camera length of 91mm (Z contrast imaging condition). Images show the AlOx layer, metallic Al islands, and GaAs substrate. 
		\textbf{(b,d,f)} Intensity profile along red lines on (a,c,e), showing width of Al hills. The lines goes from left to right red circles. 
		(b) 6 nm out of 47 nm is covered with Al hills. 13\% hill coverage.
		(d) 24 nm out of 45 nm is covered with Al hills. 53\% hill coverage. 
		(f) 35 nm out of 80 nm is covered with Al hills. 44\% hill coverage.
	}
	
	\label{figSTEM}
	
\end{figure*}

As shown on Fig. 2(a) in main text, AO seems to create metallic Al hills, $\sim2$ nm in height. To estimate how much of the GaAs surface they cover, three long range STEM scans were taken, see Fig. \ref{figSTEM}, from which an average of 1/3 coverage was found.

\subsection{EELS analysis}
To characterize the Al(Ox) in between the metallic Al hills, EELS line scans were acquired along the growth direction in 10 different regions.
From EELS scans, the material composition (eg. the presence of oxygen) along that line can be determined, as elaborated in Ref.~[42]. In the Al L-edge spectra metallic aluminum causes intensity peaks at 97 eV, while for alumina, two peaks appear at 79 eV and 98 eV and GaAs has a shoulder around 110 eV. Furthermore, aluminum and alumina can be distinguished by the presence of peaks around 540 eV in O K-spectra which is a signature of oxygen, shown in Ref.~[43].

\begin{figure*}[h]
	\includegraphics[width=0.7\linewidth]{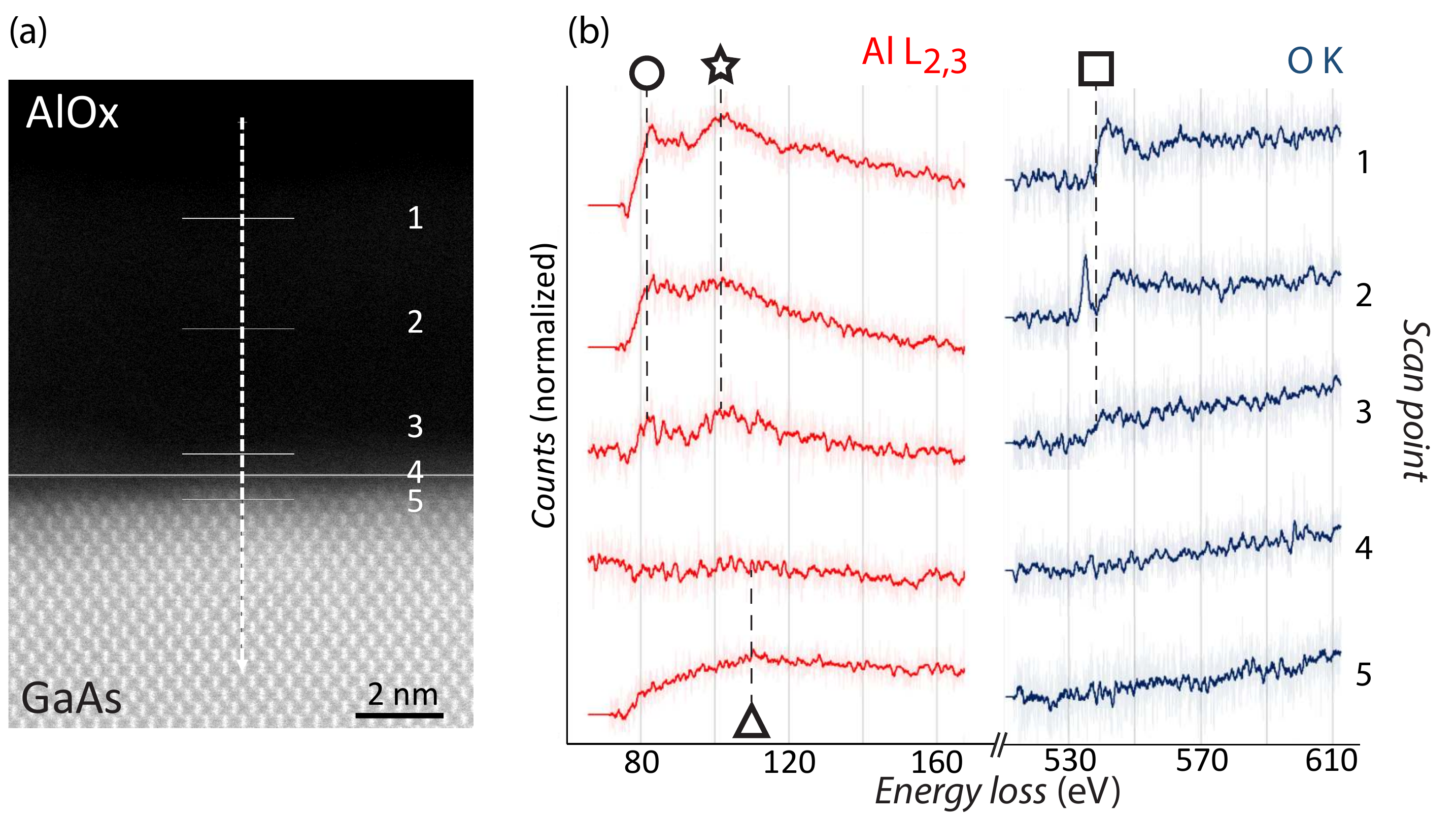}
	\caption{
		\textbf{(a)} 
		High resolution scanning transmission micrograph in high-angle annular dark-field mode focusing on the Al/GaAs interface. Electron energy loss spectroscopy (EELS) was performed at 5 different locations (marked by horizontal lines) along the growth direction.
		Points 3, 4, and 5 are spaced by a 0.5nm step size.
		\textbf{(b)}  EELS image showing L edge for Al spectra (Red) and K edge for O (Blue) from scan points 1--5, in (a). Symbols indicate composition: ($\circ$) AlOx, ($\star$) Al or AlOx, ($\triangle$) GaAs, ($\Box$) oxygen.
		The Al L2,3 edge goes from fully oxidized in points (1-2) to an oxidized state in point 3, to starting a small edge of GaAs in point 4. The O K edge shows a similar trend, with oxygen structure observed in points 1, 2 and 3 while not present for point 4.
	}
	\label{figEELS1}
\end{figure*}

\begin{figure*}[h]
	\includegraphics[width=0.76\linewidth]{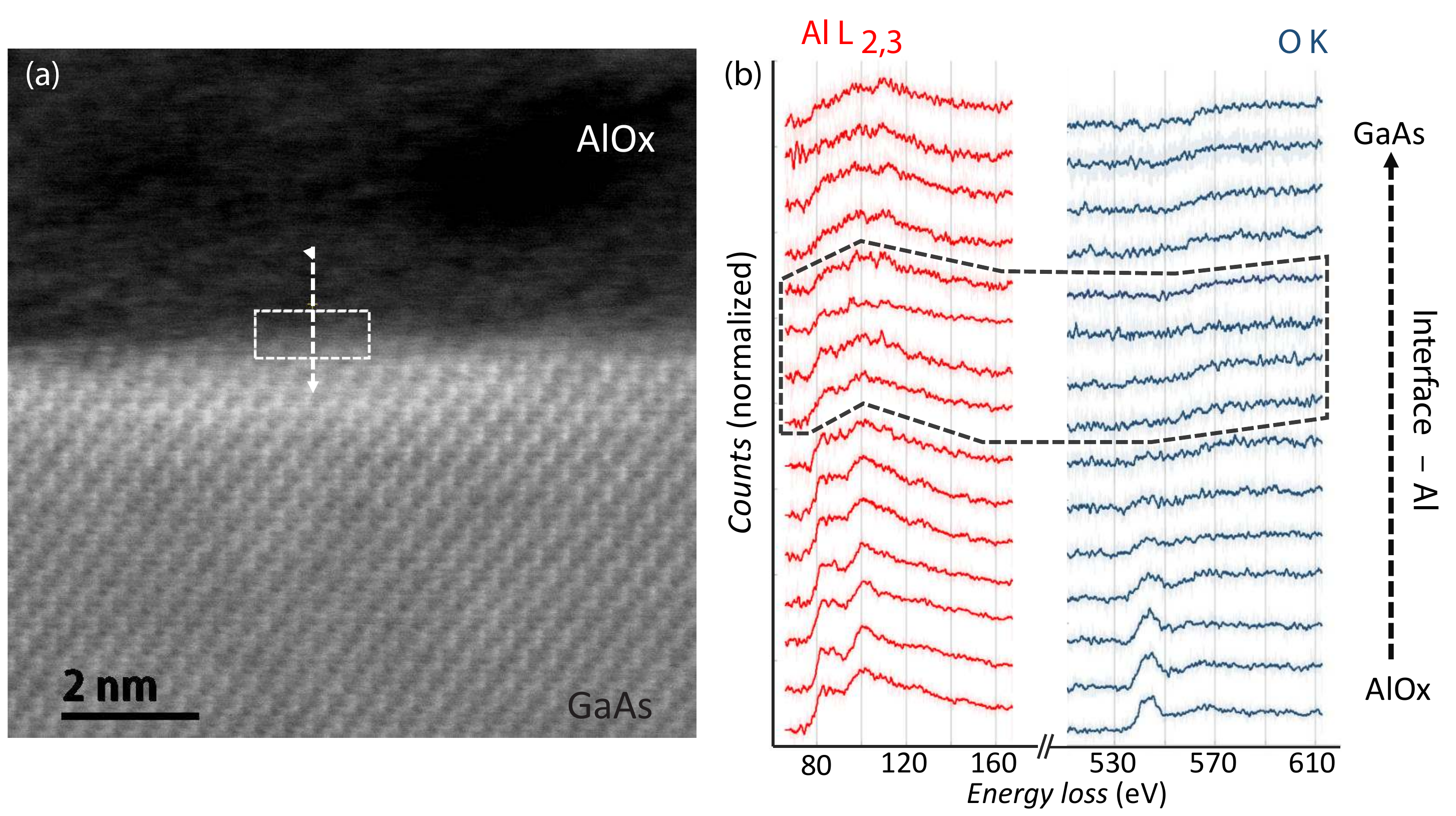}
	\caption{
		\textbf{(a)} 
		Scanning tunneling electron microscopy high-angle annular dark-field image with 37 mm camera length focusing on the Al/GaAs interface. 
		Electron energy loss spectroscopy (EELS) was performed at 15 consecutive steps, with 0.14 nm spacing along the dashed line to characterize composition.
		The dashed square marks the vertical extent of four consecutive points
		\textbf{(b)} EELS image showing L edge for Al spectra (Red) and K edge for O (Blue). The dashed polygon
		highlights the four consecutive points, regionally marked in (a), each with indications of Al without AlOx features. The bottom-most scans are from the AlOx region while the top-most are from the GaAs.
	}
	\label{figEELS2}
\end{figure*}

Out of the 10 linescans, only one, displayed on \ref{figEELS1}, indicates no traces of metallic Al at the interface.  Thus the data suggests that $\sim10\%$ of the space in between hills is fully oxidized. Examples of scans that shows oxygen-free aluminum is shown on Fig. 2(b,c) and Fig. \ref{figEELS2}.

\newpage

\hspace{1cm}

\newpage

\section{Extra scans of Al on GaAs}
\label{E}

\begin{figure*}[h]
	\includegraphics[width=\linewidth]{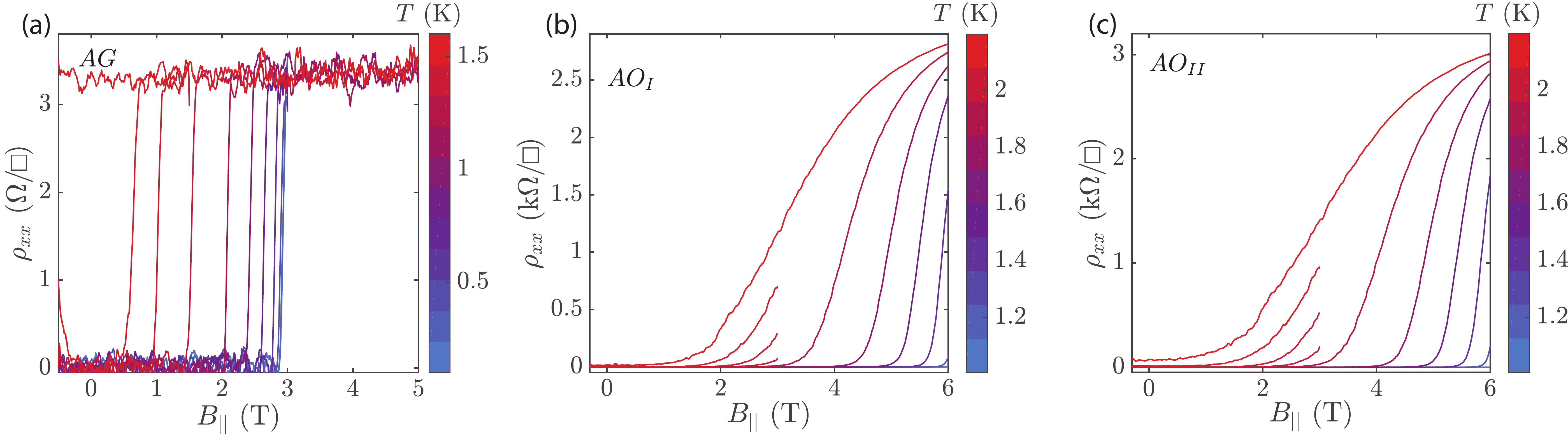}
	
	\caption{
		Sample 1 resistivity measurements as a function of $B_{||}$ at different temperatures for 
		\textbf{(a)} the as-grown $AG$ Al bar,
		\textbf{(b)} $AO_I$ Hall bar, and
		\textbf{(c)} $AO_{II}$ Hall bar.
		For (b) and (c) the lowest temperature plotted is 1.0 K, where both devices stay superconducting beyond 6 tesla.
		The data was used to extract $B_{c,||}(T)$ for Fig. 3(a) in the main text.
	}
	
	\label{figGaAsPar}
	
\end{figure*}

\begin{figure*}[h]
	\includegraphics[width=0.33\linewidth]{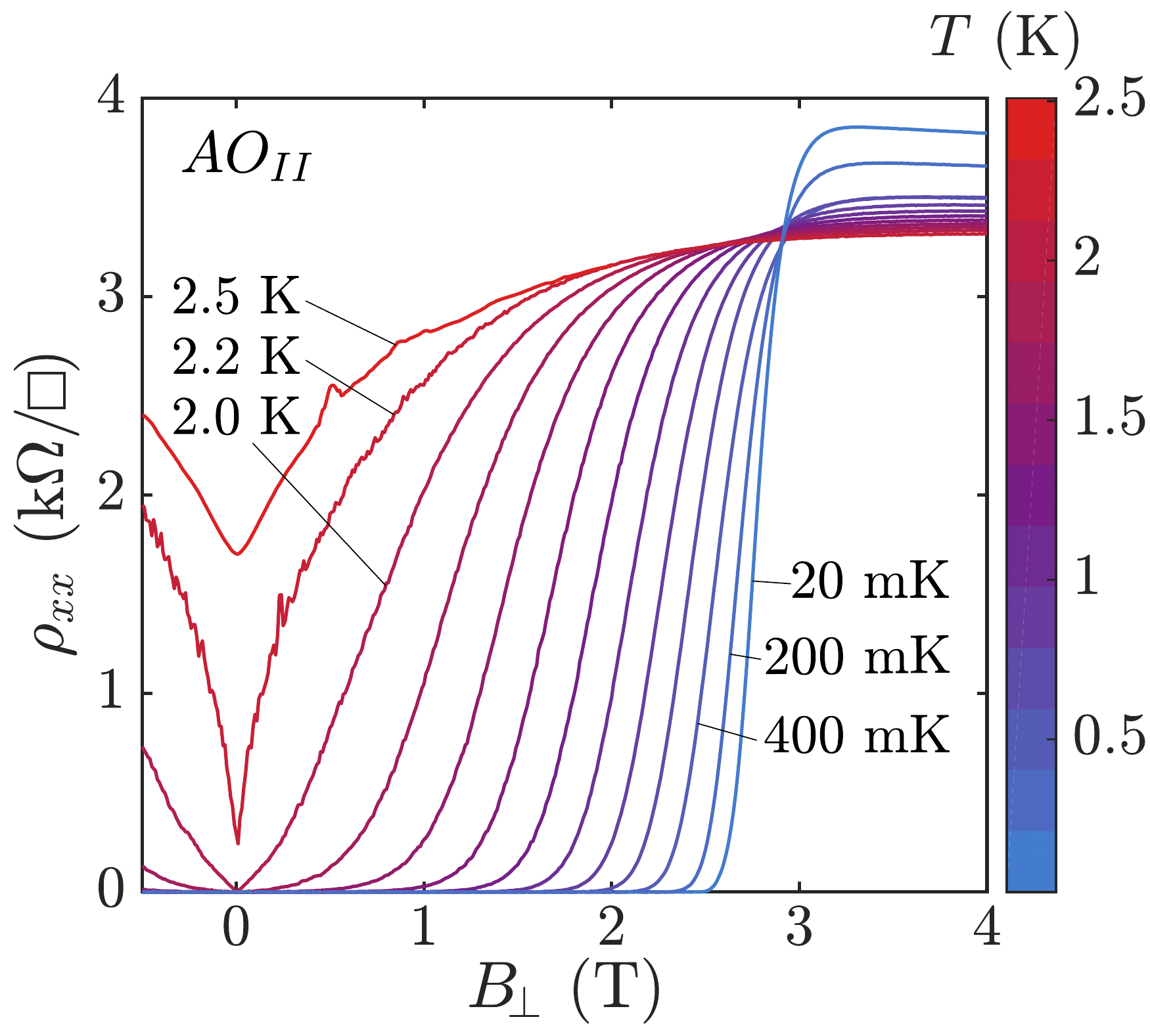}
	
	\caption{
		Resistivity measurements as a function of $B_{\perp}$ at different temperatures for the $AO_{II}$ Hall bar.	
		The data was used, together with data from Fig. 3(b), to extract $B_{c,\perp}(T)$ for Fig. 3(c) in the main text.
	}
	
	\label{figGaAsPerp}
	
\end{figure*}

\hspace{1cm}

\newpage

\hspace{1cm}

\section{Extra scans of Al on InAs 2DEG}
\label{F}

\begin{figure*}[h]
	\includegraphics[width=0.66\linewidth]{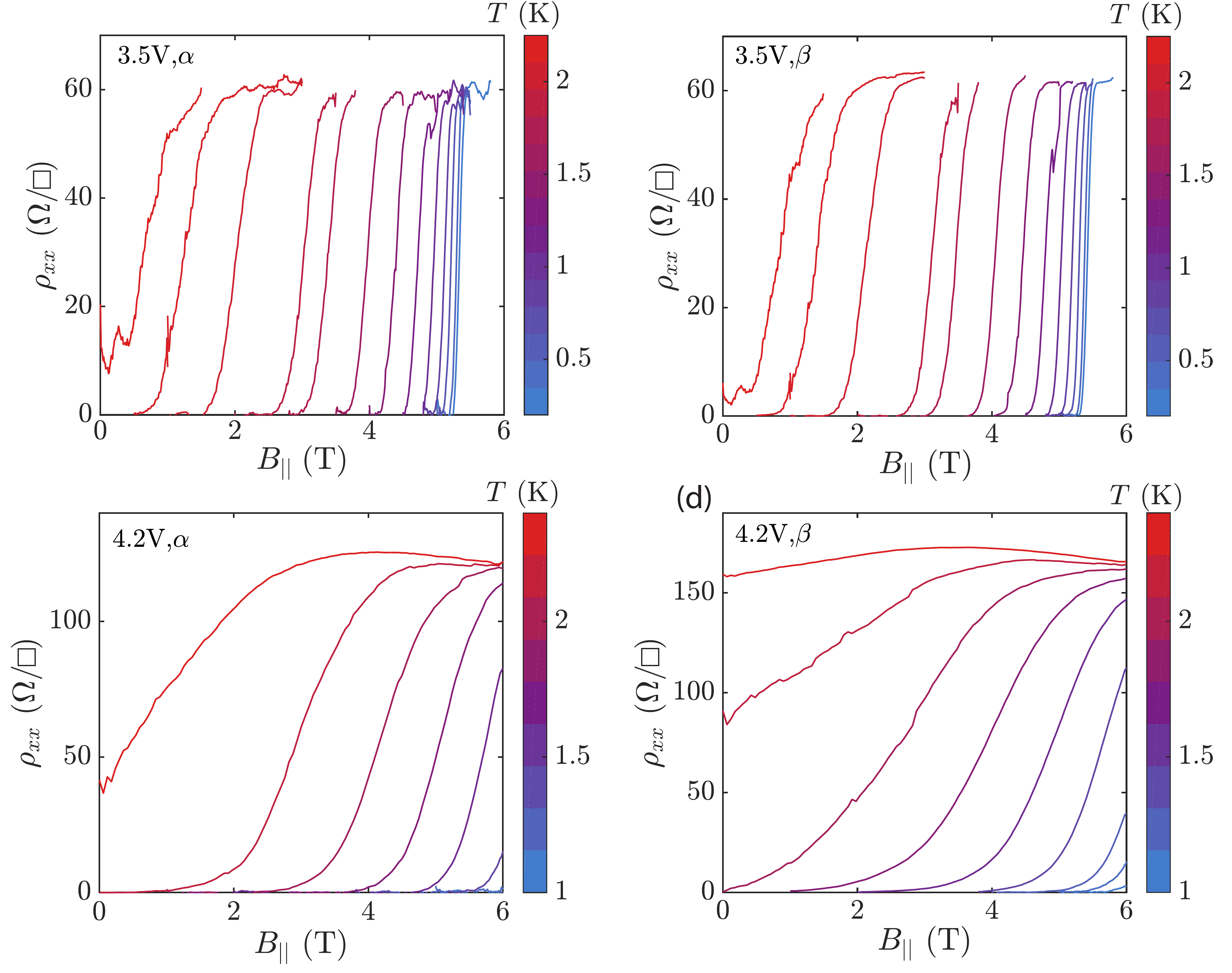}
	
	\caption{
		Resistivity measurements as a function of $B_{||}$ at different temperatures for device $\alpha$ and device $\beta$ with Hall bars anodized at 3.5 V and 4.2 V, indicated in top-left corner of each sub-figure.
		The data was used to extract $B_{c,||}(T)$ for Fig. 4(a) in the main text.
	}
	
	\label{figInAsPar}
	
\end{figure*}

\begin{figure*}[h]
	\includegraphics[width=0.33\linewidth]{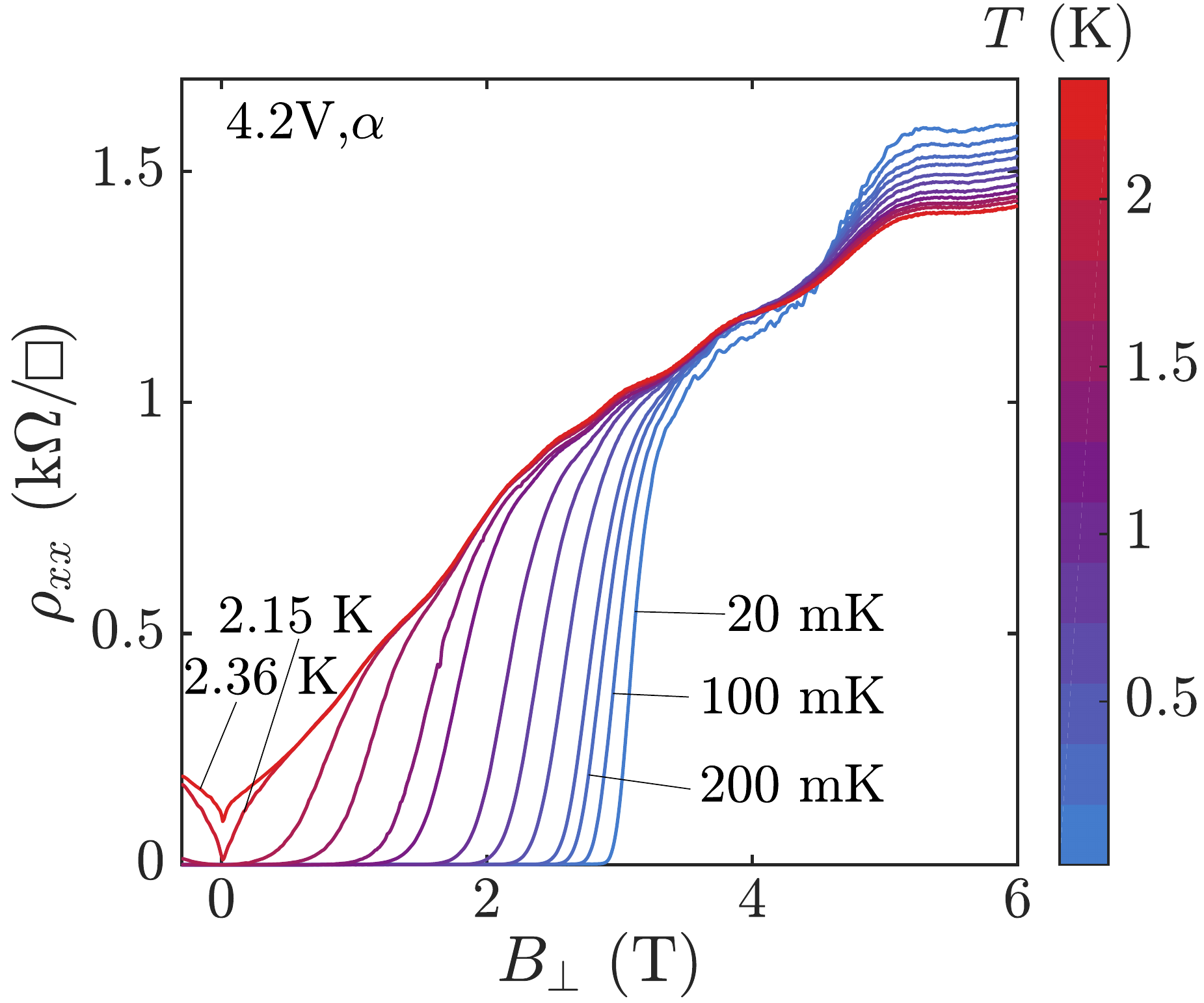}
	
	\caption{
		Resistivity measurements as a function of $B_{\perp}$ at different temperatures acquired from the Hall bar on device $\alpha$ anodized with $V_{AO}=4.2$ V.
		The data was used, together with data from Fig. 4(b), to extract $B_{c,\perp}(T)$ for Fig. 4(c) in the main text.
	}
	
	\label{figInAsPerp}
	
\end{figure*}

\newpage
\section{Hall bar fab and extra data}
\label{G}

\begin{figure*}[h]
	\includegraphics[width=0.33\linewidth]{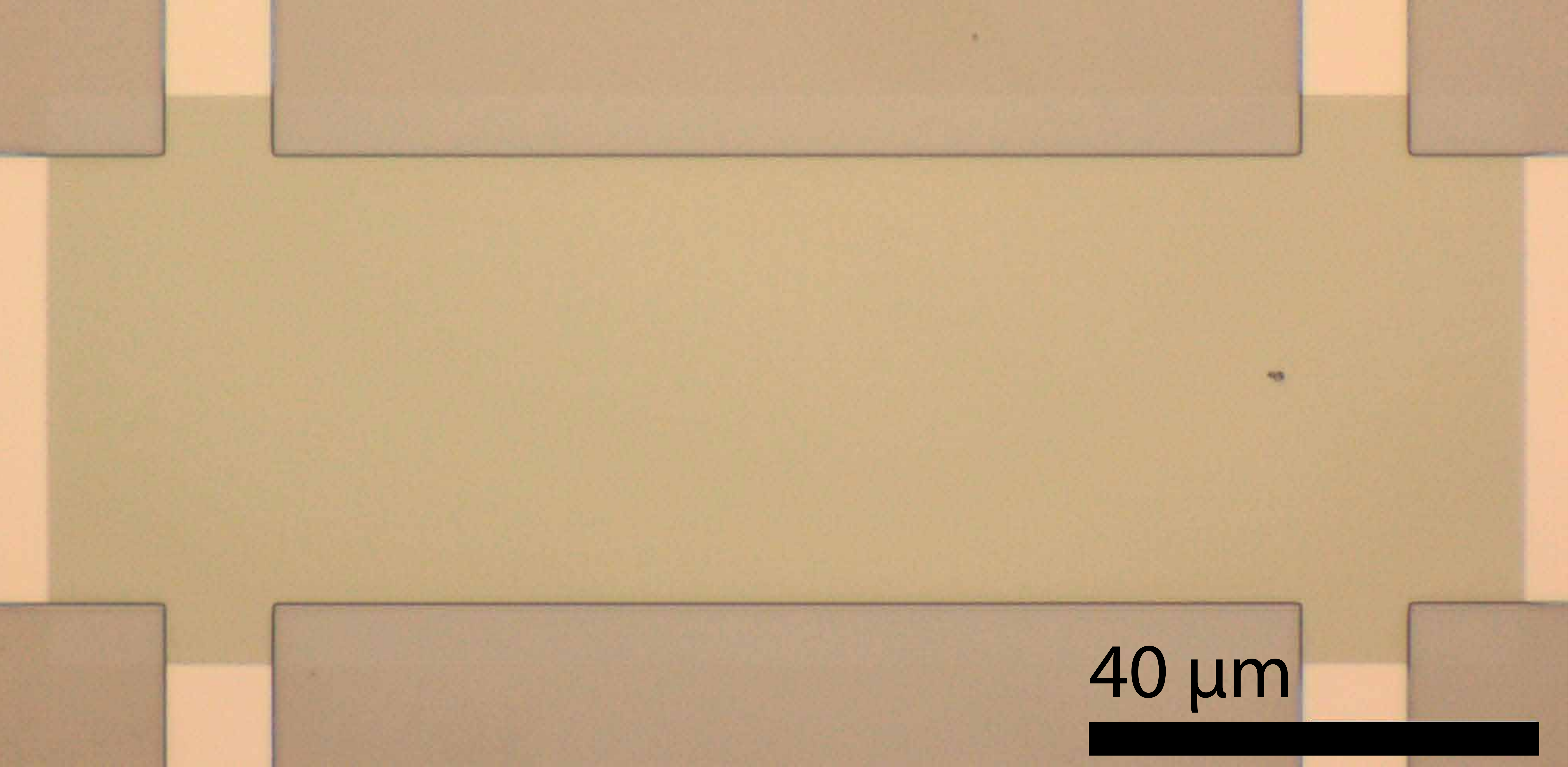}
	
	\caption{
		Optical micrograph of a Hall bar after aluminum wet etch and atomic layer deposition. The Hall bar is referred to as "Etched" on Fig. 5 in main text.
	}
	
	\label{HBEtched}
	
\end{figure*}

For comparison with anodized Hall bars, a Hall bar was fabricated using regular chemical wet etch.
After mesa etch, a window on the Hall bar, identical to the ones shown on Fig. 1(d) of main text, is patterned with electron beam lithography, see Fig \ref{HBEtched}. 
Aluminum was stripped with 5 sec in 50 $^\circ$C hot Transene Al Etchant type D, followed by 20 sec rinse in 50 $^\circ$C hot MQ water, followed by 40 sec in room temperature MQ water, and dried by blowing N$_2$. The resist is stripped and the chip is immediately loaded into atomic layer deposition (ALD). ALD and top-gates were deposited as described in Sec. \ref{C}.  \\

\begin{figure*}[h]
	\includegraphics[width=\linewidth]{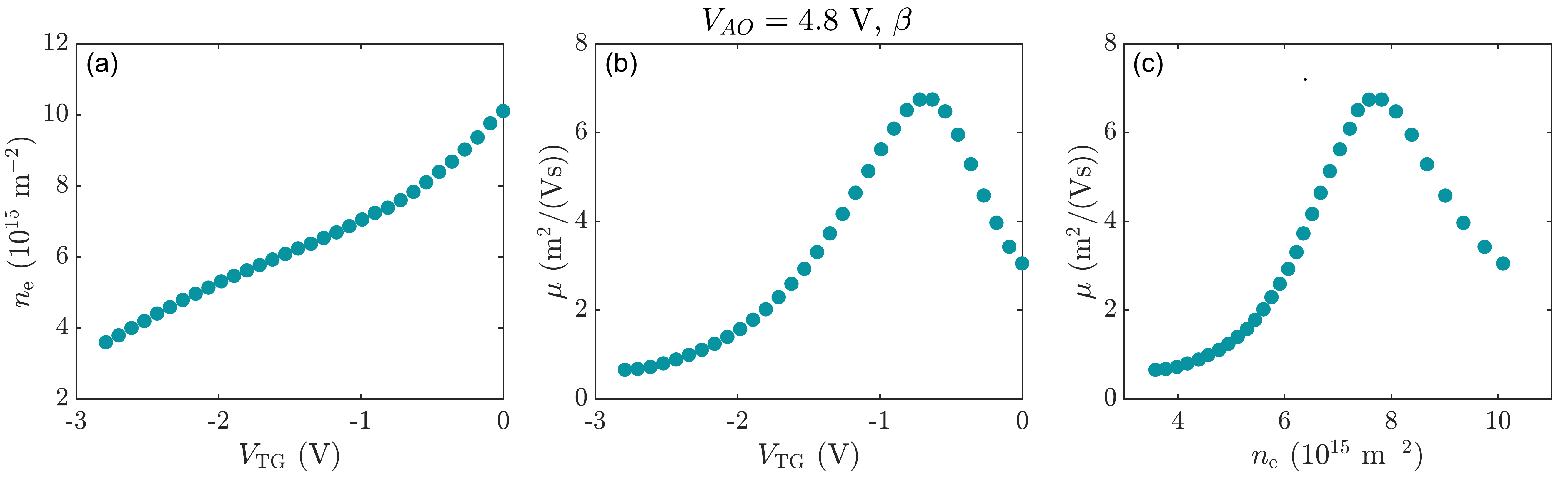}
	
	\caption{
		Dependence of Hall bar mobility and density versus top-gate voltage for the 4.8 V anodized Hall bar on device $\beta$. 
		\textbf{(a)} Density as a function of top-gate voltage.
		\textbf{(b)} Mobility as a function of top-gate voltage.
		\textbf{(c)} Parametric plot of mobility as a function of density.	
	}
	
	\label{fig70mob}
	
\end{figure*}

In addition to the anodized Hall bars mentioned in the main text, another Hall bar, also exposed to AO at 4.8~V but on device $\beta$, was fabricated and measured, see \ref{fig70mob}. It's mobility peak of $6.7\times10^4~\mathrm{cm^2/Vs}$ at $n_\mathrm{e}=7.7\times10^{11}~\mathrm{cm^2/Vs}$, came out lower than the 4.8 V Hall bar in the main text which had $8.0\times10^4~\mathrm{cm^2/Vs}$ at an equal density. 
As mentioned in the main text, the applied AO setup, elaborated in Sec. A, is not sufficient to reproduce results close to full oxidation of the Al. Here a small change in oxidation depth significantly impacts semiconductor properties after full oxidation.

\newpage
\section{Additional quantum Hall data}
\label{H}

\begin{figure*}[h]
	\includegraphics[width=\linewidth]{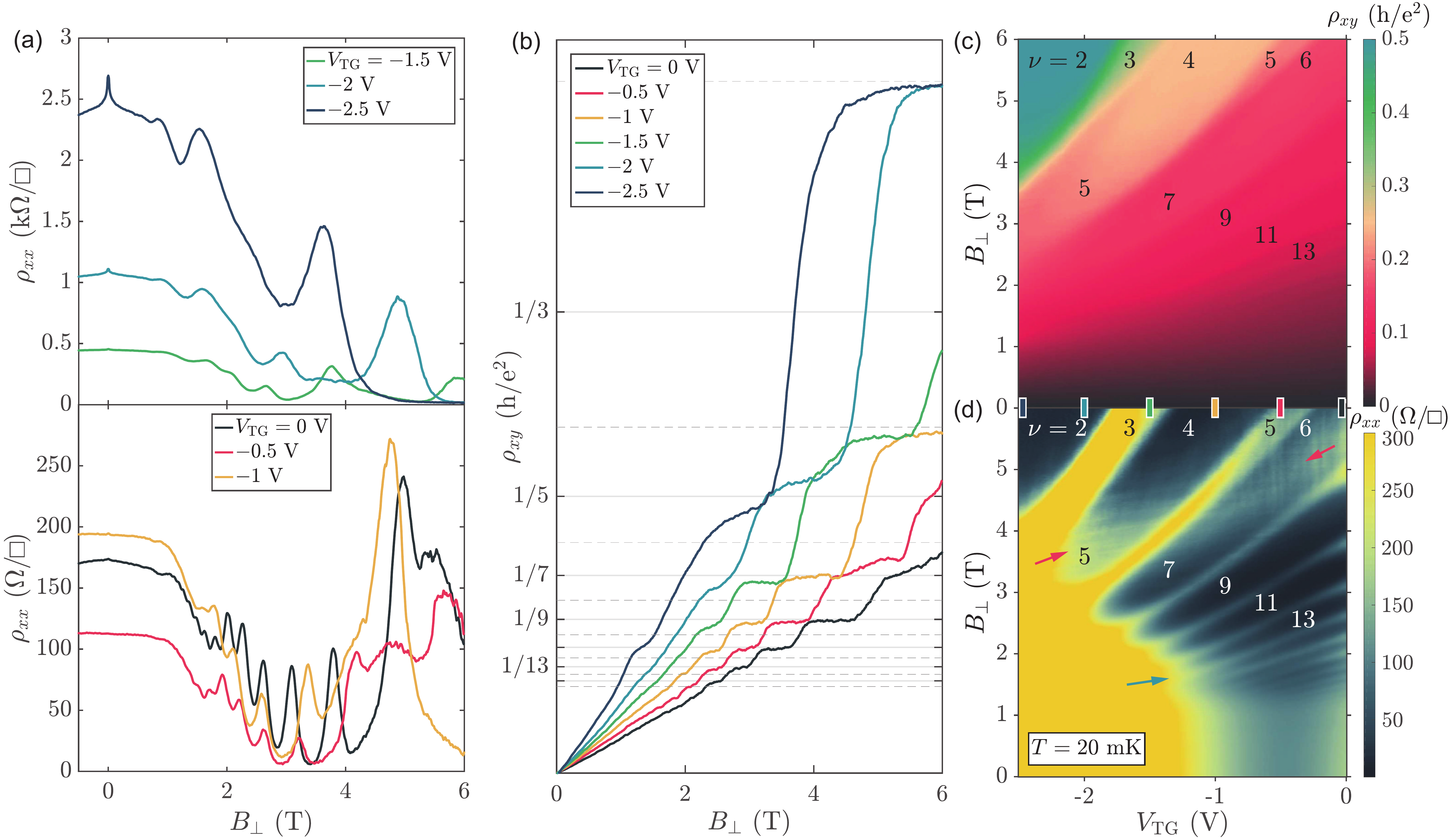}
	
	\caption{
		Quantum Hall regime at base temperature for Hall bar anodized at 4.8 V.
		\textbf{(a)} Longitudinal and \textbf{(b)} Hall resistivities as a function of $B_\perp$ for different top-gate voltages. 
		In (b) the quantum Hall resistivities for odd (solid) and even (dashed) integer filling factors marked. 
		2D map of the \textbf{(c)} Hall  and \textbf{(d)} longitudinal  resistivities as a function of $B_\perp$ and top-gate voltage, $V_\mathrm{TG}$. Extracted filling factors are indicated on the maps. 
		Arrows in (d) highlight additional resonances.
	}
	
	\label{OtherQHE}
	
\end{figure*}

The Hall bar anodized at 4.8V on device $\beta$ was also measured in high perpendicular magnetic fields, see Fig. \ref{OtherQHE}. 
Shubnikov-de Haas oscillations are observed in sub-figure (a) but $\rho_{xx}$ does not reach zero for intermediate $B_\perp$. Therefore, the plateaus in Hall resistivity, seen in sub-figure (b), deviate from the integer values. 
Weak localization is also present in $\rho_{xx}$ of this Hall bar, broader in field and taller in $\rho_{xx}$, compared to the Hall bar anodized at $4.8~\mathrm{V}$ on device $\alpha$ from main text. This indicates the 4.8~V anodized Hall bar on device $\beta$ has lower electron mean free path in agreement with the lower mobility measured in Fig. \ref{fig70mob}.

\begin{figure*}[h]
	\includegraphics[width=0.7\linewidth]{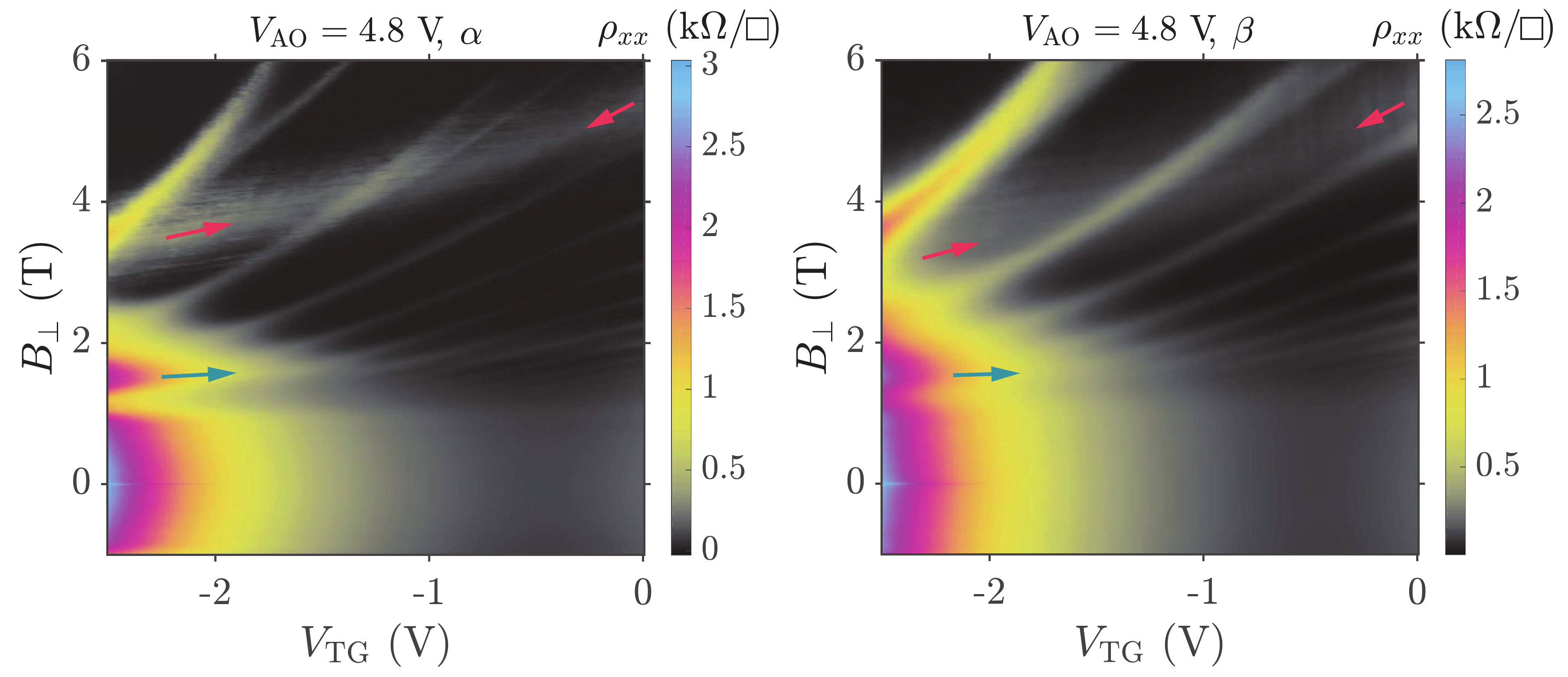}
	
	\caption{
		Longitudinal resistivity versus top-gate voltage and perpendicular field for the two Hall bars anodized at 4.8 V on device $\alpha$ and $\beta$.
		Same data as shown in Fig. 6(d) and Fig. \ref{OtherQHE}(d), but without color saturation, allowing us to see more features including the weak localization peak around zero perpendicular magnetic field and negative top-gate voltages.
		The resonances marked on Fig. 6(d) and Fig. \ref{OtherQHE}(d) are also marked with arrows here. Especially the resonances below 2 T, (blue arrows) are more striking on these plots.
	}
	
	\label{NoSatu}
	
\end{figure*}

Inspecting the $\rho_{xx}$ data from Fig.~\ref{OtherQHE}(a) and (d), we again see additional resonances (marked by arrows), similar to those observed in Fig. 6 of the main text. Both bars showing the resonances indicates that they are material or process related, not just a spurious effect on a single device. The resonances on both devices are more visible on Fig.~\ref{NoSatu}, which plots the same data as Fig. 6(d) and Fig. \ref{OtherQHE}(d), but without color saturation.

The number of edge modes, $\nu$, is extracted from Fig.~\ref{OtherQHE}(c) and added to (d). As for the 4.8~V anodized Hall bar on device $\alpha$ we observe that crossing the resonance in Fig.~\ref{OtherQHE}(d) (marked by red arrows) adds 1 to $\nu$.

\newpage
\section{Trying to pattern AO with electron beam resist}
\label{I}

A good lateral resolution ($<50$ nm) EBL mask is required to establish certain devices taking advantage of the enhanced performance of both superconductor and III/V. 

\begin{figure*}[h]
	\includegraphics[width=0.7\linewidth]{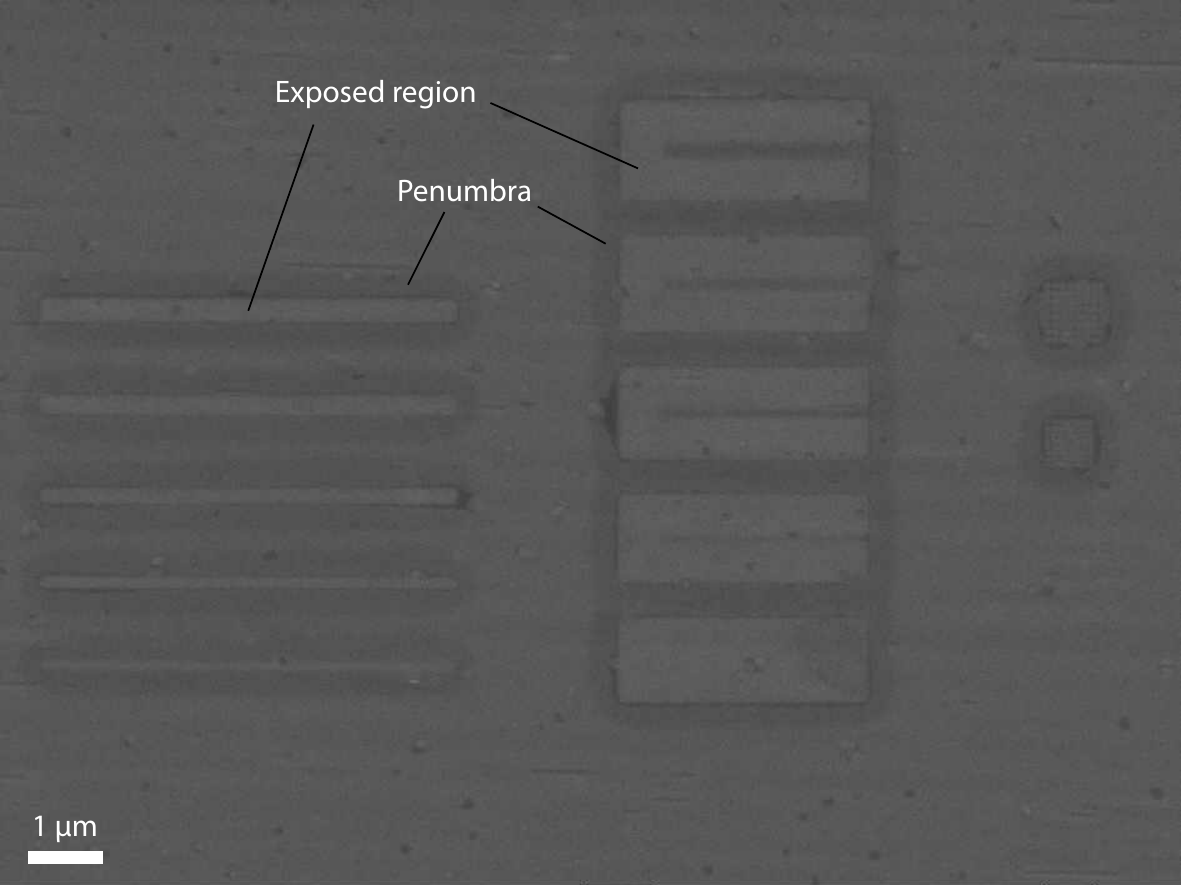}
	
	\caption{
		Scanning electron microscopy of a test patterning of AO with an electron beam lithography resist mask. The bright regions are the developed regions directly exposed by AO. The darker 'penumbra' are regions around the exposed areas, unintendedly  being affected by AO.
	}
	
	\label{Penumbra}
	
\end{figure*}

Various EBL resist masks were tested, but all of them had the same problem of a penumbra forming in under the resist around the exposed areas. The penumbra was observed with scanning electron microscopy after stripping the resist, one example is displayed in Fig.~\ref{Penumbra}. 

The size of the penumbra, ranging from 0.2 to 1 $\mu$m, was dependent on AO time and voltage, but was somewhat uniform for a given AO.
The chemical composition of the penumbra is not known, but the color change observed with scanning electron microscopy made us pursue other masking approaches.

\newpage
\section{Characterization of Semi- and superconductor properties of metal mask fab}
\label{J}
\begin{figure*}[h]
	\includegraphics[width=\linewidth]{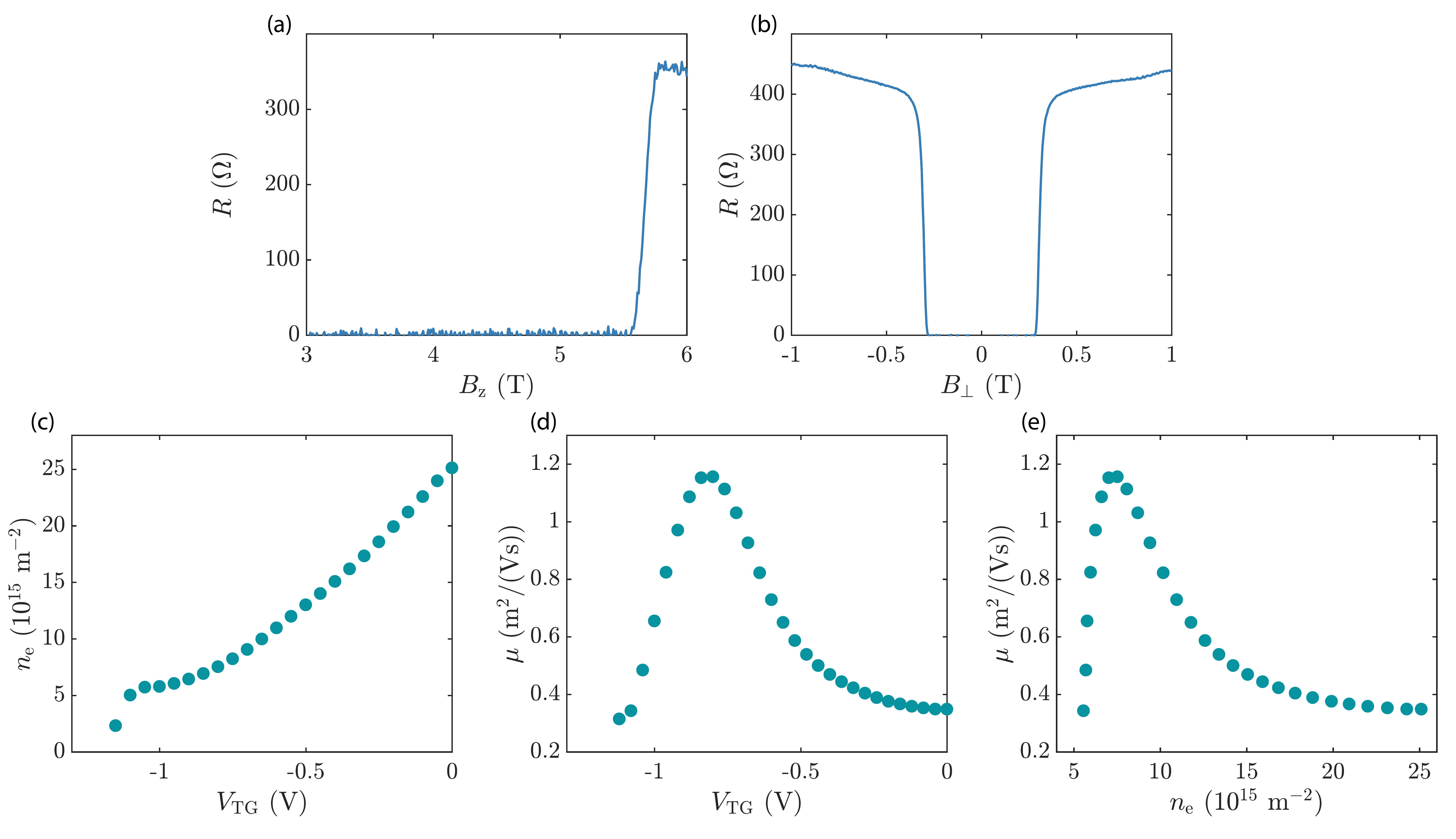}
	
	\caption{
		Characterization of superconducting bar and Hall bar after double AO with Ti mask, measured at 20~mK. From the same chip which gave the data of Fig. 7 in the main text. 
		Resistance versus \textbf{(a)} in-plane magnetic field and \textbf{(b)} out of plane magnetic field.
		\textbf{(c)} Hall bar density as a function of top-gate voltage.
		\textbf{(d)} Hall bar mobility as a function of top-gate voltage.
		\textbf{(e)} Hall bar mobility as a function of density.	
	}
	
	\label{TiChara}
	
\end{figure*}

A titanium mask was used to establish high-resolution lithography, see Sec. VII of the main text. The chip had, among other devices, two test bars. One was exposed with both AO steps, becoming semiconducting. The other would have 3 nm Ti evaporated on it in between the two AO steps, protecting the underlying Al from the second AO to keep it metallic/superconducting. The first $V_{\mathrm{AO}}=3.5~\mathrm{V}$ was chosen to resemble the 3.5 V bars on Fig. 4 in main text. This came out as intended with $B_{c,||}=5.6$ T and  $B_{c,\perp}=300$ mT, see Fig.~\ref{TiChara}(a,b). The second $V_{\mathrm{AO}}=4.8~\mathrm{V}$ was chosen to resemble the 4.8 V exposure, giving high mobility peaks, but didn't work out well. A mobility peak $\mu=1.2\times10^4\mathrm{cm^2/Vs}$ was measured at a density of $n_\mathrm{e}=7.3\times10^{11}$~$\mathrm{cm^{-2}}$, see Fig.~\ref{TiChara}(c-e), most likely due to the AO having a time dependence, see Sec. \ref{A}. A time dependence would cause the two AO to add up in depth of oxidation, thus oxidizing too deep.
As shown in Fig 5 of the main text: a too deep oxidation can degrade the mobility.

\newpage
\section{Data from other Josephson junction devices}
\label{K}
\begin{figure*}[h]
	\includegraphics[width=\linewidth]{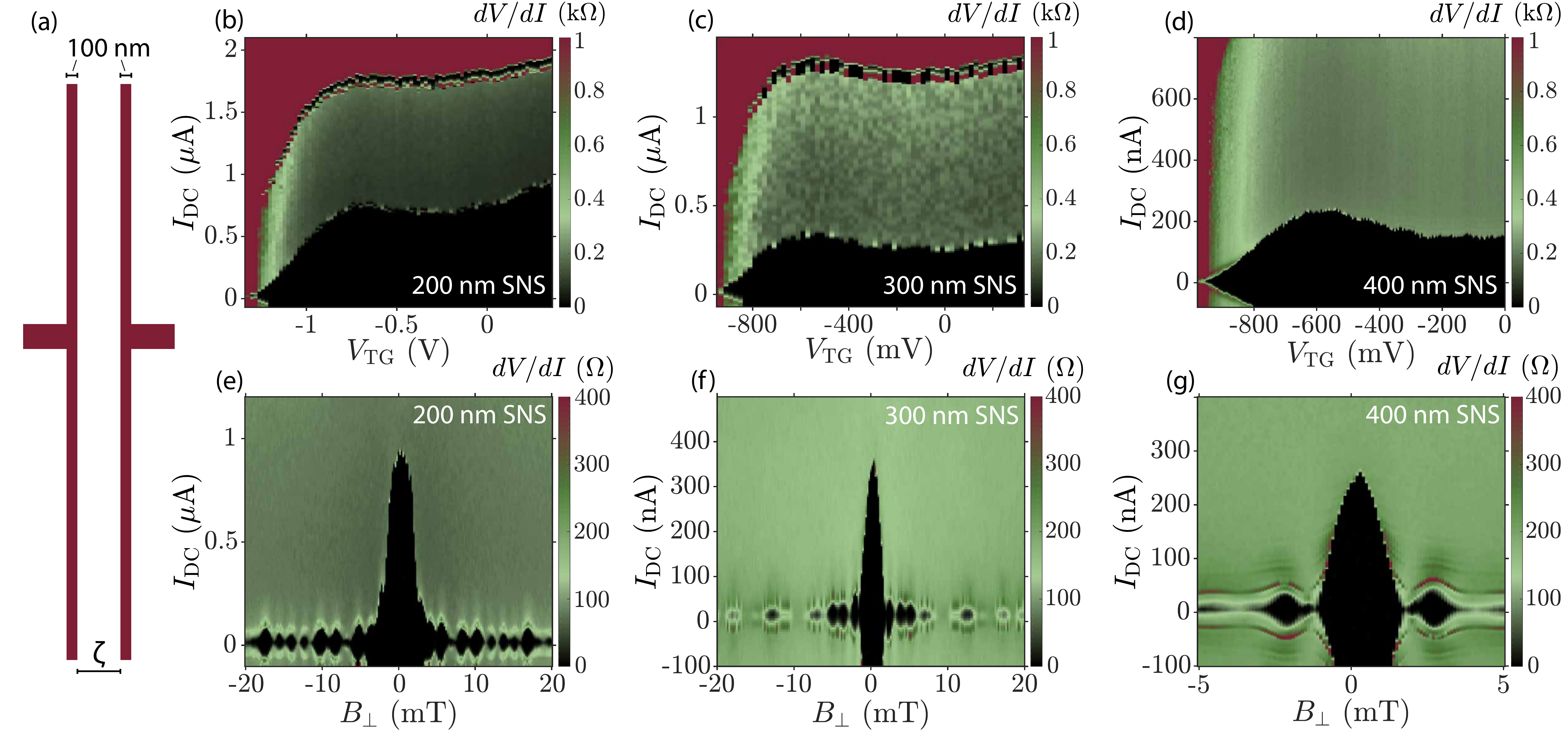}
	
	\caption{
		Data from other Josephson junctions (JJs) using the same Ti mask for AO lithography.
		\textbf{(a)} Schematic of the design shape of Al for the JJs, introducing the superconductor separation, $\zeta$.
		\textbf{(b-d)} Resistance as a function of DC current bias and top-gate voltage for devices with $\zeta=200$ nm, 300 nm and 400 nm.
		\textbf{(e-g)} Resistance as a function of DC current bias and perpendicular magnetic field, for devices with $\zeta=200$ nm, 300 nm, and 400 nm, marked in right corner of each sub-figure.	
	}
	
	\label{OtherSNS}
	
\end{figure*}

Besides the Josephson junction (JJ) reported on Fig. 7 in the main text, other JJs were fabricated with increased separations, $\zeta$, between the superconducting leads, indicated on Fig. \ref{OtherSNS}(a). The rest of the figure shows supercurrent gate-ability and Fraunhofer oscillations in JJs with $\zeta=200$ nm, 300 nm and 400 nm. We have thus shown that thin elongated structures reproducibly can be manufactured with AO using a metal mask. 

The gate voltages used are shown in the table below:

\begin{center}
	
	\begin{tabular}{c | c | c}
		\hline
		$\zeta$ & $V_w$ (V) & $V_\mathrm{TG}$ (V) in Fraunhofer scan \\\hline
		200 nm & -1.2 & -0.56 \\
		300 nm & -1.50 & 0.36 \\		
		400 nm & -0.70 & -0.55 \\
		\hline
	\end{tabular}
	
\end{center}

\newpage
\section{Resistance oscillations in 100 \lowercase{nm} Josephson junction}
\label{L}
\begin{figure*}[h]
	\includegraphics[width=1\linewidth]{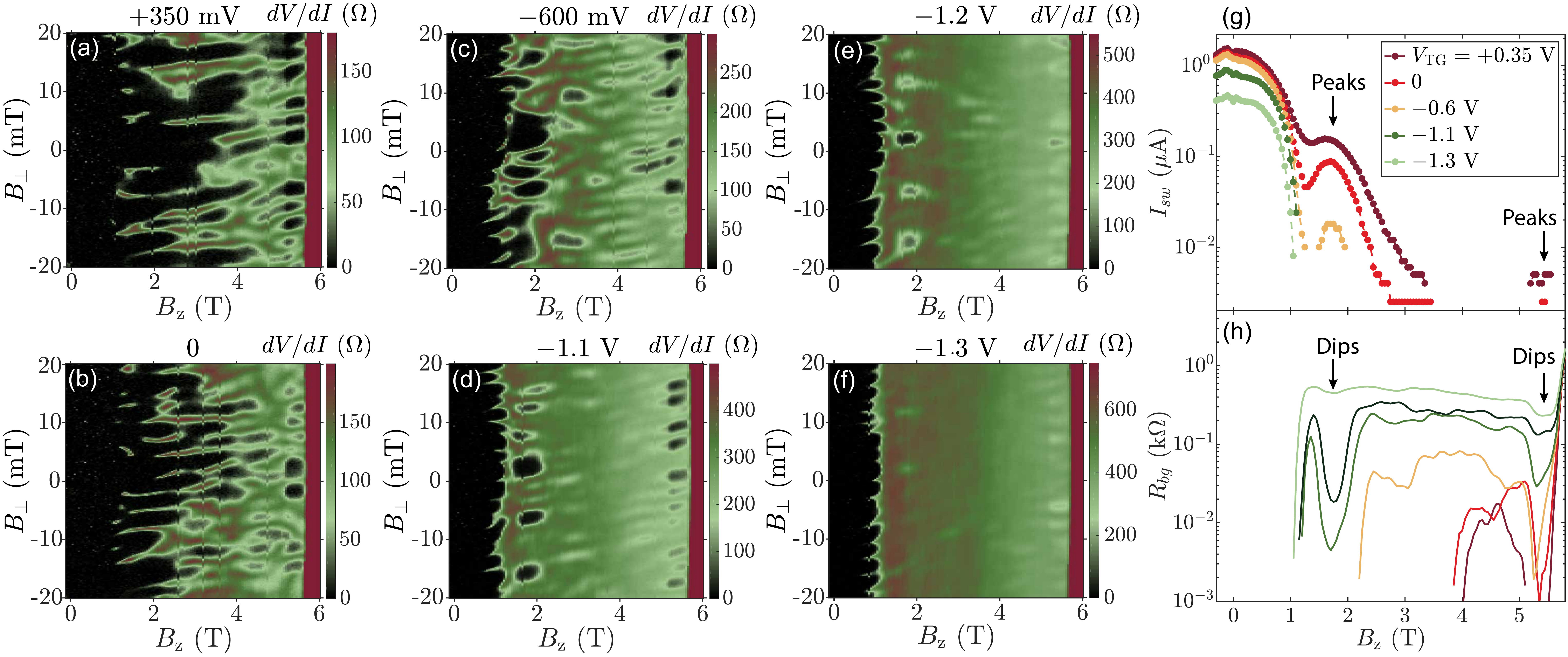}
	
	\caption{
		Analysis of supercurrent oscillations as a function of magnetic field and top-gate voltage. Differential resistance is plotted as a function of $B_z$ and $B_\perp$ for different $V_\mathrm{TG}$, indicated above the saturated color maps: 
		\textbf{(a)} $+350$ mV,
		\textbf{(b)} $0$,
		\textbf{(c)} $-600$ mV,
		\textbf{(d)} $-1.1$ V,
		\textbf{(e)} $-1.2$ V and
		\textbf{(f)} $-1.3$ V.
		\textbf{(g)} Replica of Fig. 7(g) from main text. Oscillation and re-emergence of switching current, $I_\mathrm{sw}$ (see main text) vs $B_{z}$ for several top-gate voltages, $V_\mathrm{TG}$.
		\textbf{(h)} Back ground resistance versus $B_z$ for different $V_\mathrm{TG}$.
	}
	
	\label{Rbg}
	
\end{figure*}

For the $\zeta=100$ nm JJ reported in Fig. 7 of the main text, the oscillatory behavior of superconductivity versus an in-plane field $B_z$(perpendicular to current direction) was observed. Oscillations can also be seen in the JJs resistive state. To display this, maps of differential resistance, $dV/dI$, as a function of $B_z$ and $B_\perp$, were taken at different top-gate voltages, see Fig. \ref{Rbg} (a-f). In each map, the minimum $dV/dI$ for every value of $B_z$ was found. We call it the background resistance $R_{bg}(B_z,V_{TG})$. 

Plotting $R_{bg}$ below the switching current $I_{sw}$ data of Fig. 6(f), shows that dips in $R_{bg}$ happens at the same $B_z$ giving peaks in $I_{sw}$ independent of $V_\mathrm{TG}$, see Fig. \ref{Rbg} (g-h). This indicates a lock of density dependence.

\end{document}